\documentclass[3p, preprint]{elsarticle}
\usepackage[utf8]{inputenc}

\usepackage{amsmath,amsbsy,amsthm,amssymb}
\usepackage{multirow,multicol,tabularx,booktabs}
\usepackage{graphicx,placeins,color,url}
\usepackage{bbm,bm, mathrsfs}
\usepackage[ruled]{algorithm2e}
\usepackage[shortlabels]{enumitem}
\usepackage[bookmarks=false]{hyperref}
\usepackage{capt-of}

\newtheorem{conj}{Conjecture}
\newtheorem{thm}[conj]{\bf Theorem}

\newtheorem{rem}[conj]{\bf Remark}

\newtheorem{example}[conj]{\bf Example}

\def\bar{\overline}
\def\to{\rightarrow}

\def\Dc{\mbox{$\mathcal D$}}

\def\Fc{\mbox{$\mathcal F$}}

\def\Ic{{\mathcal I}}

\def\Kc{{\mathcal K}}
\def\Lc{\mbox{$\mathcal L$}}

\def\Nc{\mbox{$\mathcal N$}}

\def\Sc{{\mathcal S}}
\def\Tc{\mbox{$\mathcal T$}}
\def\Uc{\mbox{$\mathcal U$}}
\def\Vc{\mbox{$\mathcal V$}}

\def\Zc{\mbox{$\mathcal Z$}}

\def\Hb{{\mathbb H}}

\def\Nb{{\mathbb N}}

\def\Rb{{\mathbb R}}
\def\Wb{{\mathbb W}}

\def\EE{ {\rm I} \kern-.15em {\rm E} }
\def\PP{ {\rm I} \kern-.15em {\rm P} }

\def\t{ {\bf t}}
\def\u{ {\bf u}}

\def\x{ {\bf x}}
\def\y{ {\bf y}}
\def\z{ {\bf z}}

\def\X{ {\bf X}}

\def\Z{ {\bf Z}}
\def\psibm{ {\bm \psi}}

\def\mds{\medskip}
\def \1{\mathbbm{1} }
\def\eps{\varepsilon}

\journal{Computational Statistics and Data Analysis}

\begin{document}

\begin{frontmatter}

    \title{A classification point-of-view about conditional Kendall’s tau}
    
    \author[crest]{Alexis Derumigny}
    \ead{alexis.derumigny@ensae.fr}
    
    \author[crest]{Jean-David Fermanian}
    \ead{jean-david.fermanian@ensae.fr}
    \address[crest]{CREST-ENSAE, 5, avenue Henry Le Chatelier,
    91764 Palaiseau Cedex, France.}
    
    \date{\today}
    
    \begin{abstract}
        We show how the problem of estimating conditional Kendall's tau can be rewritten as a classification task. Conditional Kendall's tau is a conditional dependence parameter that is a characteristic of a given pair of random variables. The goal is to predict whether the pair is concordant (value of $1$) or discordant (value of $-1$) conditionally on some covariates. We prove the consistency and the asymptotic normality of a family of penalized approximate maximum likelihood estimators, including the equivalent of the logit and probit regressions in our framework. Then, we detail specific algorithms adapting usual machine learning techniques, including nearest neighbors, decision trees, random forests and neural networks, to the setting of the estimation of conditional Kendall's tau. Finite sample properties of these estimators and their sensitivities to each component of the data-generating process are assessed in a simulation study. Finally, we apply all these estimators to a dataset of European stock indices.
    \end{abstract}


\end{frontmatter}


\section{Introduction}

Beside linear correlations, most dependence measures between two random variables are functions of the underlying copula only: Spearman's rho, Kendall's tau, Blomqvist's beta, Gini's measure of association, etc. As a consequence, they are independent of the corresponding margins. This is seen as a positive point. See Joe~\cite{joeBook2015}, Nelsen~\cite{nelsen2007introduction}, for instance, and, as a reminder, some basic definitions in~\ref{reminder_copulas}. Such measures are well-known and widely used by practitioners. When some covariates are available, natural extensions of these tools can be defined, providing so-called ``conditional'' measures of dependence. In theory, it is sufficient to replace copulas by conditional copulas to obtain the ``conditional version'' of any dependence measure. Surprisingly, this simple and fruitful idea has not yet been widely used in the literature. 
Nonetheless, in a series of papers, Gijbels et al.~\cite{gijbels2011conditional,gijbels2011Scandinav,gijbels2012EJS,gijbels2015EJS} have popularized this approach, with a focus on conditional Kendall's tau and Spearman's rho. 
Note that conditional dependence measures have been invoked in different frameworks, often without any explicit link with conditional copulas: truncated data (Tsai~\cite{tsai1990truncation}, e.g.), multivariate dynamic models (Jondeau and Rockinger~\cite{jondeau2006}, Almeida and Czado~\cite{almeida2017}, among others), vine structures (So and Yeung~\cite{soyeung2014regularized}), etc.

\mds

Now, let us introduce our key dependence measure: for each $\z \in \Rb^p$, the conditional Kendall's tau of a bivariate random vector $\X := (X_1,X_2)$ given some covariates $\Z=\z$ may be defined as
\begin{align*}
    \tau_{1,2|\Z=\z}
    &= \PP \big( (X_{2,1} - X_{1,1})(X_{2,2} - X_{1,2}) > 0
    \big| \Z_1 = \Z_2 = \z \big) - \PP \big( (X_{2,1} - X_{1,1})(X_{2,2} - X_{1,2}) < 0 \big| \Z_1 = \Z_2 = \z \big) ,
\end{align*}
where $\X_1=(X_{1,1},X_{1,2})$ and $\X_2=(X_{2,1},X_{2,2})$ are two independent versions of $\X$. 
To simplify, we will assume that the law of $\X$ given $\Z=\z$ is continuous w.r.t. the Lebesgue measure, for every $\z$. 
This implies 
\begin{align*}
    \tau_{1,2|\Z=\z}
    &= 2 \, \PP \big( (X_{2,1} - X_{1,1})(X_{2,2} - X_{1,2}) > 0
    \big| \Z_1 = \Z_2 = \z \big) - 1.
\end{align*}

A conditional Kendall's tau belongs to the interval $[-1 , 1]$ and reflects a positive ($\tau_{1,2|\Z=\z} > 0$) or negative ($\tau_{1,2|\Z=\z} < 0$) dependence between $X_1$ and $X_2$, given $\Z=\z$. Contrary to correlations, it has the advantage of being always defined, even if some $X_k$, $k=1,2$, has no finite second moments (when it follows a Cauchy distribution, for example).

\mds

Some estimators of conditional Kendall's tau have already been proposed in the literature, either as a by-product of the estimation of conditional copulas - see Gijbels et al.~\cite{gijbels2011conditional} and Fermanian and Lopez \cite{fermanian2015single} - or directly, as in Derumigny and Fermanian \cite{derumigny2018kendall, derumigny2018kernelBased}. Nonetheless, to the best of our knowledge, nobody has yet noticed the relationship between conditional Kendall's tau and classification methods.

\mds

Let us explain this simple idea. 
Denote $W:= 2 \times \1 \{  (X_{2,1} - X_{1,1})(X_{2,2} - X_{1,2}) > 0 \} - 1$ and
$$\PP \big( (X_{2,1} - X_{1,1})(X_{2,2} - X_{1,2}) > 0
\big| \Z_1 = \Z_2 = \z \big)
= \PP \big( W = 1 \big| \Z_1 = \Z_2 = \z \big)=: p(\z).$$
Actually, the prediction of concordance/discordance among pairs of observations $(\X_1,\X_2)$ given $\Z$
can be seen as a classification task of such pairs.
If a model is able to evaluate the conditional probability of observing concordant pairs of observations,
then it is able to evaluate conditional Kendall's tau, and the former quantity is one of the outputs of most classification techniques. 
Therefore, most classifiers can potentially be invoked (for example linear classifiers, decision trees, random forests, neural networks and so on~\cite{friedman2001elements}), but applied here to pairs of observations. 

\mds

Indeed, for every $1 \leq i,j \leq n$, $i \neq  j $, define $W_{(i,j)}$ as 
\begin{equation}
    W_{(i,j)} := 2 \times \1 \{  (X_{j,1} - X_{i,1})(X_{j,2} - X_{i,2}) > 0 \} - 1 =
    \begin{cases}
        1 & \text{if } (i,j) \text{ is a concordant pair,} \\
        -1  & \text{if } (i,j) \text{ is a discordant pair}.
    \end{cases}
    \label{eq:def_W_ij} 
\end{equation}
A classification technique will allocate a given couple $(i,j)$ into one of the two categories $\{1,-1\}$ (or ``concordant versus discordant'', equivalently), with a certain probability, given the value of the common covariate~$\Z$.

\mds

Section \ref{section:reg_approach} introduces a general regression-type approach for the estimation of conditional Kendall's tau. Some asymptotic results of consistency and asymptotic normality are stated. In Section \ref{section:algorithms}, we explain how some machine learning techniques can be adapted to deal with our particular framework, and we detail the corresponding algorithms. A small simulation study compares the small-sample properties of all these algorithms in Section \ref{section:simulation}. In Section \ref{section:applications_financial}, these techniques are applied to European stock market data. We evaluate to what extent the dependence between pairs of European stock indices may change with respect to different covariates. All proofs have been postponed into appendices.

\section{Regression-type approach}
\label{section:reg_approach}

Typically, a regression-type model based on conditional Kendall's tau may be written as
\begin{equation}
    \tau_{1,2|\Z=\z}=g_0(\z,\beta^*), \;\; \forall \z\in \Zc\subset \Rb^p,
    \label{model:param_cond_tau}
\end{equation}
for some finite dimensional parameter $\beta^*\in \Rb^{p'}$ and some function $g_0$.
As a particular case, a single-index approach would be
\begin{equation}
\tau_{1,2|\Z=\z}=g(\psibm(\z)^T\beta^*) , \;\; \forall \z\in \Zc ,
\label{model:cond_tau_Z}
\end{equation}
where $\psibm: \Rb^p \rightarrow \Rb^{p'}$ is known, and $g$ may be known (parametric model) or not (semi-parametric model), as in~\cite{derumigny2018kendall}. 
In this section, we propose an inference procedure of $\beta^*$ under~(\ref{model:cond_tau_Z}) when the link function $g$ is analytically known. This procedure will be based on the signs of pairs only, and not on the specific values of the vectors $\X_i$.
Then, since inference will be based on the observations of $W\in \{1,-1\}$, our model belongs to the family of limited-dependent variable methods. One difficulty will arise from the pointwise conditioning events $\Z_i=\Z_j=\z$, that will necessitate localization techniques.
Actually, we will consider couples of observations $\X_i$ and $\X_j$ for which the associate covariates are close to a given value $\z$. Indeed, the relationship~(\ref{model:cond_tau_Z}) does not define the dependence levels between every couple $(\X_i,\Z_i)$ and $(\X_j,\Z_j)$, $i\neq j$, but only between those that share the same covariate. If the variables $\Z$ were discrete, we would consider a subset of couples such that $\Z_i=\Z_j$. In our case of continuous variables $\Z$ (see below), the latter event does not occur almost surely, and some smoothing/localization techniques have to be invoked.  


\mds

Let $K$ be a $p$-dimensional kernel and $(h_n)$ be a bandwidth sequence. The bandwidth will simply be denoted by $h$ and we set $K_h(\z)=K(\z/h)/h^p$.
The log-likelihood associated to the observation $(W_{(i,j)},\Z_i,\Z_j)$ given $\Z_i=\Z_j=\z$ is
$$    \ell_\beta(W_{(i,j)},\z)
    := \bigg(\frac{1+W_{(i,j)}}{2}\bigg)
    \log \PP_\beta \bigg(W_{(i,j)} = 1 \Big| \Z_i=\Z_j =\z \bigg) + \bigg(\frac{1-W_{(i,j)}}{2} \bigg)
    \log\PP_\beta \bigg(W_{(i,j)} = -1 \Big| \Z_i=\Z_j = \z \bigg).$$
In practice, when the underlying law of $\Z$ is continuous, there is virtually no couple for which $\Z_i=\Z_j$.
Therefore, we will consider a localized ``approximated'' log-likelihood, based on $(W_{(i,j)},\Z_i,\Z_j)$ for all pairs $(i,j)$, $i\neq j$. It will be defined as the double sum
$$ L_n(\beta):=  \frac{1}{n(n-1)}\sum_{i,j;i\neq j} K_h(\Z_i - \Z_j) \ell_\beta (W_{(i,j)},\tilde{\Z}_{i,j}),$$
for any choice of $\tilde{\Z}_{i,j}$ that belongs to a neighborhood of $\Z_i$ or $\Z_j$. 
We will assume that $K$ is a compactly supported $p$-dimensional kernel of order $m\geq 2$.

\mds

The most obvious choices would be to select $\tilde{\Z}_{i,j}$ among $ \{\Z_i,\Z_j,(\Z_i+\Z_j)/2\}$. 
Here, we propose
\begin{eqnarray*}
\lefteqn{    L_n(\beta)
    := \frac{1}{n(n-1)}\sum_{i,j;i\neq j} K_h(\Z_i - \Z_j) \ell_\beta (W_{(i,j)},\Z_i) }\\
    &=& \frac{1}{n(n-1)}\sum_{i,j;i\neq j} K_h(\Z_i - \Z_j) \left\{
    \bigg(\frac{1+W_{(i,j)}}{2}\bigg)  \log \bigg( \frac{1}{2} +
    \frac{1}{2} g ( \psibm( \Z_{i})^T \beta) \bigg) \right. \\
    &+& \left. \bigg(\frac{1-W_{(i,j)}}{2}\bigg)\log
    \bigg( \frac{1}{2} - \frac{1}{2} g ( \psibm( \Z_i)^T \beta) \bigg) \right\},
\end{eqnarray*}
under~(\ref{model:cond_tau_Z}).
We can therefore derive an estimator of $\beta^*$ based on the maximization of the latter function, with a $\ell_1$ penalty (Lasso-type estimator), as
\begin{equation}
    \hat \beta
    := \arg \max_{\beta \in \Rb^{p'}} L_n(\beta) - \lambda_n |\beta|_1,
    \label{estime_beta}
\end{equation}
where $\lambda_n$ is a tuning parameter to be chosen.
Note that $L_n(\beta)$ is not really a likelihood function since the observations $(W_{(i,j)},\Z_i,\Z_j)$ for every couple $(i,j)$, $i\neq j$, are not mutually independent. 

\mds

If $K\geq 0$, the objective function is a concave function of $\beta$ if it satisfies
\begin{equation}
    \delta g''(t)(1+\delta g)(t) \leq (g')^2(t), \; \forall t,
    \label{convex_g}
\end{equation}
for $\delta\in\{1,-1\}$.
When $\beta \mapsto L_n(\beta)$ is concave, the penalized criterion above is concave too and the calculation of $\hat\beta$ can be led in practical terms through convex optimization routines, even with a large number of regressors ($p' \gg 1$). Since this will be our framework, we will show that~(\ref{convex_g}) holds for some usual classification techniques. When it is not the case, we have to rely on other optimization schemes and to avoid considering too many regressors.

\mds

Moreover, note that, when $g$ is odd (i.e. $g(-t)=-g(t)$),
the estimator simply becomes
\begin{equation}
    \hat \beta
    := \arg \max_{\beta \in \Rb^{p'}}
     \frac{1}{n(n-1)} \sum_{i,j;i\neq j} K_h(\Z_i - \Z_j)
    \log \left( \frac{1}{2} +
    \frac{1}{2} g ( W_{(i,j)}\psibm( \Z_{i})^T \beta) \right)  - \lambda |\beta|_1.
    \label{simplified_criterion}
\end{equation}

\mds

Implementation of an algorithm to solve problem (\ref{estime_beta}) or its simplified version (\ref{simplified_criterion}) may seem difficult due to the non-differentiability of the $l_1$ norm. Nevertheless, as in the case of the ordinary Lasso, it can be solved in a very efficient way using the Alternative Direction Method of Multipliers (ADMM) for general $l_1$ minimization, following \cite[Section 6.1]{boyd2011distributed}. More precisely, assume $L_n(\beta)$ is a concave and differentiable function of $\beta$ (this is the case in both Examples~\ref{example:logit} and~\ref{example:probit}).
Then the optimization task~(\ref{estime_beta}) can be rewritten as finding the solution~$(\x,\z) \in \Rb^{2p'}$ of
\begin{equation}
    \left\{ \begin{aligned}
        \text{minimize } f(\x) + g(\z)\\ 
        \text{subject to } \x - \z = 0,
    \end{aligned} \right.
    \label{optimization_pb_rewritten}
\end{equation}
where $f(\x) := - L_n(\x)$ and $g(\z) := \lambda_n |\z|_1$.
The solution is given by iterating the following algorithm, denoting by $\u \in \Rb^{p'}$ the dual variable of the problem~(\ref{optimization_pb_rewritten}) and by~$\rho > 0$ the step size (similarly to the usual gradient descent algorithm)
\begin{align*}
    &\x^{k+1}
    := \arg \min_\x \big( f(\x) + (\rho/2) ||\x-\z^k+\u^k||_2^2 \big), \\
    &\z^{k+1}
    := S_{\lambda_n / \rho} (\x^{k+1} + \u^k ), \\
    &\u^{k+1} := \u^k + \x^{k+1} - \z^{k+1},
\end{align*}
where for any $\kappa > 0$, $S_\kappa$ is the element-wise soft thresholding operator, i.e. for each component $S_\kappa(a) := (1-\kappa/|a|)_+ \times a$, for $a \neq 0$, and $S_\kappa(0) := 0$.
Note that we have reduced the non-differentiable problem~(\ref{estime_beta}) into a sequence of differentiable optimization steps for $\x$, and the computation of the proximal operator $S_\kappa$ for the $\z$-updates. We refer to \cite{parikh2014proximal} for a detailed presentation about proximal operators and their use in optimization.
ADMM can also be adapted for large-scale data, using standard libraries and frameworks for parallel computing such as MPI, MapReduce and Hadoop, see \cite{boyd2011distributed} for more details about implementation of such methods.

\mds

\begin{example}[Logit]
    If we choose the Fisher transform $g(t)=(e^t-1)/(e^t+1)$, then $g$ is odd and the optimization program becomes:
    \begin{align*}
        \hat \beta
        := \arg \max_{\beta \in \Rb^{p'}}
        \frac{1}{n(n-1)} \sum_{i,j;i\neq j} K_h(\Z_i - \Z_j)
        \log \left( logit( W_{(i,j)}\psibm( \Z_{i})^T \beta ) \right)  - \lambda |\beta|_1,
    \end{align*}
    where the so-called logit link function is defined by $logit(x) = e^x / (1 + e^x)$.
    Therefore $\hat \beta$ can be seen as the maximizer of the log-likelihood of a weighted logistic regression with independent realizations of an explained variable $W_{(i,j)}$, given some explanatory variables $\Z_i$. On a practical side, when $K\geq 0$, the $\beta$-criterion is concave. This allows to use the existing software and optimization routines of logistic regression without many changes.
    \label{example:logit}
\end{example}

\begin{example}[Probit]
    Similarly, choosing $g(t) = 2 \Phi(t)-1$, where $\Phi$ denotes the cdf of the standard normal distribution, yieds the equivalent of a (weighted) probit regression. Indeed, this function $g$ is odd, ~(\ref{simplified_criterion}) applies in this case and our criterion in~(\ref{estime_beta}) is concave wrt $\beta$. 
    \label{example:probit}
\end{example}



\mds

Let us assume that a family of models or some statistical procedure allow the calculation of the functional link $g(\epsilon\psi(\z)^T\beta)$ and then $p(\z)$, for any $\z$, $\epsilon\in\{-1,1\}$ and any given value $\beta$:
Logit, Probit, regression trees, neural networks, etc. Then, we can estimate the ``true'' parameter $\beta^*$ by $\hat\beta$, as given by~(\ref{estime_beta}), in practical terms. 

\mds

Now, we state the
asymptotic properties of $\hat \beta$, under the assumption that $\beta\mapsto L_n(\beta)$ is concave. To this goal, we introduce some notations.

\mds

For any $\x$ and $\y \in \Rb^p$, denote
$$p(\x,\y):=\PP_{\beta^*}( (X_{2,1} - X_{1,1})(X_{2,2} - X_{1,2}) > 0 | \Z_1=\x,\Z_2=\y) .$$
The latter expectations are calculated when the underlying parameter is assumed to be the true value $\beta^*$.
Note that $p(\x):=p(\x,\x)$ and $2p(\z)-1=\tau_{1,2|\Z=\z}$.
Moreover, for any $\x,\y$ and $\z \in \Rb^p$, set
$$    p(\x,\y,\z) :=\PP_{\beta^*}( (X_{2,1} - X_{1,1})(X_{2,2} - X_{1,2}) > 0, 
    (X_{3,1} - X_{1,1})(X_{3,2} - X_{1,2}) > 0| \Z_1=\x,\Z_2=\y,\Z_3=\z) .$$
This is the conditional probability that $\X_1$, $\X_2$ and $\X_3$ are concordant, given their respective covariates. Denote, for any $\beta \in \Rb^{p'}$,
$$  \phi(\x,\y,\beta) := p(\x,\y)  \log \left( q(\x,\beta) \right)+ (1- p(\x,\y) ) \log  \left( 1-q(\x,\beta) \right) , \;
q(\x,\beta):= 1/2 + g ( \psibm( \x)^T \beta)/2.$$
Note that $q(\z,\beta^*)=p(\z)$.
Finally, for any real function $f$ and $\eps>0$, denote by $f_\eps$ the function $x\mapsto \sup_{t,\;|x-t|<\eps} |f(t)|$.

\mds

{\it Regularity assumption R0:}
The density $f_\Z$ of $\Z$ is assumed to be $m$-times continuously differentiable. Moreover, the functions $\phi(\x,\cdot,\beta)$ and $q(\cdot,\beta)$ are continuous for any $\x \in \Zc$ and any $\beta \in \Rb^{p'}$. To simplify, $(\x,\y,\z)\mapsto p(\x,\y,\z)$ will be continuous on $\Zc^3$.

\mds

\begin{thm}
    \label{consistency_beta}
    Under R0,~(\ref{cond_consis_1}) and~(\ref{cond_consis_2}) in~\ref{proof_consistency}, if $\lambda_n \rightarrow 
    \lambda_\infty$ and $n^2h^p\rightarrow \infty$ when $n\rightarrow \infty$, if the true model is given by~(\ref{model:cond_tau_Z}) and $\beta\mapsto L_n(\beta)$ is concave, then 
    the solution $\hat\beta$ of~(\ref{estime_beta}) tends in probability towards $ \beta^{**}:=\arg\max_\beta L_{\infty}(\beta) - \lambda_{\infty} |\beta |_1$, where
    $$ L_{\infty}(\beta):= \int \phi(\z,\z,\beta) f^2_{\Z}(\z) \, d\z.$$
\end{thm}

In particular, when $\lambda_\infty=0$, the estimator $\hat\beta$ tends to $\arg\max_\beta L_{\infty}(\beta) =\beta^*$, because 
$\phi(\z,\z,\beta)$ is the expected log-likelihood associated to $W_{(1,2)}$ given $\Z_1=\Z_2=\z$. Thus, for every $\z$, the latter quantity is maximal when $\beta=\beta^*$.

\begin{thm}
    \label{AN_beta}
    Under the conditions of Theorem~\ref{consistency_beta} and some additional conditions of regularity in~\ref{proof_AN} (notably (\ref{cond_reg_AN_phi}), (\ref{cond_reg_fZphi}), (\ref{cond_reg_H1H2}) and (\ref{cond_reg_abc})), if $n^{1/2}\lambda_n\rightarrow \mu$ and $nh^p\rightarrow \infty$ when $n\rightarrow \infty$, then
    $n^{1/2}(\hat\beta-\beta^*)$ weakly tends to 
    $$ \u^*=\arg\min_{\u \in \Rb^p} 
    \Wb(\beta^*) \u+ \frac{1}{2}\u^T \Hb(\beta^*) \u-
     \mu \sum_{k;\beta_k^*=0} |u_k| - \mu\sum_{k;\beta_k^* \neq 0} sign(\beta_k^*) u_k ,$$
    where $\Wb(\beta^*) \sim \Nc(0_p,\Sigma_{\beta^*})$,
    $ \Sigma_{\beta^*} = \int \partial_\beta\phi(\z,\z,\beta^*)  \partial_{\beta}\phi(\z,\z,\beta^*)^T f^3_\Z(\z) \, d\z$
     and $$\Hb(\beta^*) = \int \partial^2_{\beta,\beta^T} \phi(\z,\z,\beta^*) f^2_{\Z}(\z) \, d\z.$$
\end{thm}

\begin{rem}
    All the previous results and those of the next sections are based on the kernel-weighted log-likelihood criterion $L_n(\beta)$, and then on the choice of the bandwidth $h$.
    We have not tried to find an ``optimal'' smoothing parameter $h$. This task is outside the scope of this paper and is left for further research. 
    Instead, we have preferred to rely on the usual rule-of-thumb (Scott~\cite{scott1992}), even if, strictly speaking, it is relevant only for kernel estimators of densities. Nonetheless, we have not empirically found an ``excessive sensitivity'' of our simulation results w.r.t. $h$.
\end{rem}

\section{Classification algorithms and conditional Kendall's tau}
\label{section:algorithms}

In the latter section, we have studied a localized likelihood procedure to estimate $\beta^*$ under~(\ref{model:cond_tau_Z}), when we can explicitly write (and code) the link function $g$.  
This may be seen as a restrictive approach, because it is far from obvious to guess the right functional form of $g$.  
To improve the level of flexibility of our conditional Kendall's tau model, 
we recall the estimation of $\tau_{1,2|\z}$ is similar to the evaluation of 
$\PP \big( W_{(1,2)}=1 \big| \Z_1 = \Z_2 = \z \big)$, i.e. the probability $p(\z)$ of classifying the couple $(1,2)$ into one of two categories (concordant or discordant), given a common value $\z$ of their covariates. 
Formally, the answer of such a question can be directly yielded by some classification algorithms. This is the topic of this section.
Therefore, instead of estimating an assumed parametric model by penalization, as in~(\ref{estime_beta}), a classification algorithm will ``automatically'' evaluate $p(\z)$ by $\hat p(\z)$. 
An estimator of the conditional Kendall's tau will simply be $\hat \tau_{1,2 |\Z=\z}:= 2 \hat p(\z) - 1$.

\mds

Now, we show how different classification algorithms can be used and adapted to the estimation of $\tau_{1,2|\Z=\z}$ in practice.
The first step is to transform the dataset
$\Dc = (X_{i,1}, X_{i,2}, \Z_i)_{i = 1, \dots, n}
\in \left( \Rb^{2+p}\right)$, called the \emph{initial dataset}, into an object $\tilde \Dc$, that will be called the \emph{dataset of pairs} (see Algorithm \ref{algo:creation_dataset_D_tilde}). Each element of this dataset of pairs is indexed by an integer $k\in \{1,\ldots, n(n-1)/2 \}$, which corresponds to any (unordered) pair $(i,j)$, $i\neq j$, of observations in the initial dataset.

\mds

For any pair of observations, we compute the associated covariate $\tilde Z_k$ which is just the average of the two covariates $\Z_i$ and $\Z_j$ (contrary to Section~\ref{section:reg_approach} where we have chosen $\Z_i$). Note that we want $\Z_i$ and $\Z_j$ to be close to each other, so that the pair $(i,j)$ is relevant. This means that a weight variable $V_k$ is defined for any pair. It is related to the proximity between $\Z_i$ and $\Z_j$. Obviously, if $V_k = 0$ then the corresponding pair is not kept, finally. This selection induces also a computational benefit, by reducing the size of the dataset and the computation time. For example, suppose that $n = 4 000$. Then, up to around $8 \times 10^6$ possible pairs can be constructed but only a small group of them (around $10^4$ or $10^5$ pairs, typically) will be relevant. The others are pairs for which the covariates are considered too far apart. Note that, in order to increase the proportion of $k$ such that the weight $V_k$ is zero, it is sufficient to use compactly supported kernels. For instance, for any arbitrary $p$-dimensional kernel $K$, we can consider $\tilde K(\z):= \gamma K(\z) \1\{\| \z \|_\infty \leq 1\}$, with some normalizing constant $\gamma$ so that $\int \tilde K=1$.

\begin{algorithm}[htb]
\label{algo:creation_dataset_D_tilde}
\SetAlgoLined
    \vspace{0.1cm}
    \KwIn{Initial dataset $\Dc = (X_{i,1}, X_{i,2}, \Z_i)_{i = 1, \dots, n}
    \in \left( \Rb^{2+p}\right)^n$ ;}
    $k \leftarrow 0$ \;
    \For{$i \leftarrow 1$ \KwTo $(n-1)$} {
        \For{$j \leftarrow (i+1)$ \KwTo $n$} {
            $\tilde \Z_k \leftarrow (\Z_i + \Z_j) / 2$ \;
            $W_k \leftarrow W_{(i,j)}$ as defined in Equation (\ref{eq:def_W_ij}) \;
            $V_k \leftarrow K_h(\Z_i - \Z_j)$ \;
            $k \leftarrow k+1$ ;
        }
    }
    Define $\Kc := \{k: V_k > 0\}$ \;
    \KwOut{A dataset of pairs $\tilde \Dc := (W_k, \tilde \Z_k, V_k)_{k \in \Kc}
    \in \big( \{-1, 1\} \times \Rb^p \times \Rb_+ \big)^{n(n-1)/2}$.}
\caption{Algorithm for creating the dataset of pairs from the initial dataset.}
\end{algorithm}

\subsection{The case of probit and logit classifiers}
\label{subsection:logit_probit_classifiers}

\mds

With the new dataset $\tilde \Dc$, we can virtually apply any classification method to predict the concordance value $W_k$ of the pair $k$, given the covariate $\tilde \Z_k$ and the weight $V_k$.
The logit and probit models yield some of the oldest and easiest methods in classification. They have straightforward adapted versions in our case: see Algorithm \ref{algo:CKT_logit_probit}. 
These weighted penalized GLM procedures are estimated using the R package \texttt{ordinalNet}~\cite{wurm2017regularized}. Note that we are still estimating $\tau_{1,2|\Z=\z}$ under the parametric model given by~(\ref{model:cond_tau_Z}).
The tuning parameter~$\lambda$ can be chosen using a generalization of Algorithm 2 in Derumigny and Fermanian~\cite{derumigny2018kendall}. The chosen $\lambda$ is the one which minimizes the cross-validation criteria 
$$CV(\lambda) := \sum_{k=1}^N d\left(
\z \mapsto \hat \tau_{1,2 |\Z = \z}^{(k)}
\, ; \,
\z \mapsto g\big( \psibm(\z)^T \hat\beta^{(\lambda, -k)} \big) \right),$$
where $d(\cdot \, ; \, \cdot)$ is a distance on a space of bounded functions of $\z$, for example the distance generated by the $L_2$ norm, $\hat \tau_{1,2 |\Z = \cdot}^{(k)}$ is an estimator of Kendall's tau using the dataset $\Dc_k$, $\hat\beta^{(\lambda, -k)}$ is estimated on the dataset $\Dc \backslash \Dc_k$ using the tuning parameter $\lambda$, and the initial dataset $\Dc$ has been separated at random in $N$ subsets $\Dc_1, \dots, \Dc_N$ of equal size.

\mds

\begin{algorithm}[ht]
\label{algo:CKT_logit_probit}
\SetAlgoLined
    \vspace{0.1cm}
    \KwIn{A dataset of pairs $\tilde \Dc := (W_k, \tilde \Z_k, V_k)_{k \in \Kc}$}
    \KwIn{A point $\z \in \Zc$, a function $\psibm$ and a penalty level $\lambda$;}
    Compute the usual weighted penalized logit (resp. probit) estimator $\hat \beta$ on the dataset
    $(W_k, \psibm(\tilde \Z_k), V_k)_{k \in \Kc}$ with a tuning parameter $\lambda$ \;
    \KwOut{An estimator $\hat \tau_{1,2|\Z = \z}
    := (e^{\psibm(\z)^T \hat \beta} - 1) / (e^{\psibm(\z)^T \hat \beta} + 1)$ \\ 
    (resp. $\hat \tau_{1,2|\Z = \z}
    := 2 \Phi(\psibm(\z)^T \hat \beta) - 1$).}
\caption{Estimation of the conditional Kendall's tau $\tau_{1,2|\Z = \z}$ using a logit (resp. probit) regression.}
\end{algorithm}

\subsection{Decision trees and random forests}
\label{subsection:trees_random_forests}

\mds

Now, let us discuss how partition-based methods can be used for the estimation of the conditional Kendall's tau.
Strictly speaking, such techniques are parametric: the relationship~(\ref{model:param_cond_tau})  implicitly applies, but for some complex untractable function $g$. And the parameter $\beta^*$ is related to some covariate thresholds, typically.
Nonetheless, a classical decision tree can be directly trained on the weighted dataset $\tilde \Dc$. We use the R package \texttt{tree} by Ripley \cite{ripley2018tree}, following Breiman et al. \cite{Breiman1984}.
Therefore, the application of decision trees to our framework is straightforward, and do not require any special adaptation, contrary to random forests. And the tree procedure allows the calculation of the probability of observing a concordant pair, given any common value of $\Z$.

\mds

In a classical classification setting, random forests are techniques of aggregation of decision trees that are built on a subset of samples and subsets of variables. More precisely, a typical random forest algorithm is the following: sample 80\% of the rows of the dataset (without replacement), and 80\% of the explanatory variables; estimate a tree on this, and repeat this procedure a certain number of times, with different sub-samples every time.
In our framework, it is not clear at which level subsampling should take place.

\mds

The easiest solution would be to directly plug-in the dataset of pairs $\tilde \Dc$ into a classical random forest algorithm, but it does not obviously lead to the best solution. For comparison, we detail this solution in Algorithm~\ref{algo:CKT_randomForests_bad}. We propose now an improvement on Algorithm~\ref{algo:CKT_randomForests_bad}.
Indeed, noting that aggregation of trees is useless if all trees are identical, it seems that the more variability in the input of the trees, the better. Following this idea, we have noticed that the observations in the dataset of pairs are not independent. Influence of this lack of independence is discussed in a general setting in Section~\ref{subsection:influence_independence}. For example, the pair $(1,2)$ is usually not independent of the pair $(1,3)$, because they both share the first observation $(X_{1,1}, X_{1,2}, \Z_1)$. Therefore, to increase the diversity of inputs in the different trees, we suggest to lead a first sampling $\Sc_j$ on the initial dataset, and then to build a dataset of pairs on the sampled observations $\Dc_j := (X_{i,1}, X_{i,2}, \Z_i)_{i \in \Sc_j}$ (see Algorithm \ref{algo:CKT_randomForests}).
As a matter of fact, if for example the first observation does not belong to the sample $\Sc_j$, then the dataset $\Dc_j$ and the estimated tree $\Tc_j$ become both independent of this first observation $(X_{1,1}, X_{1,2}, \Z_1)$.
This independence property makes the trees less dependent, and significantly improves the performance in our results compared to the original Algorithm~\ref{algo:CKT_randomForests_bad}.

\begin{algorithm}[htb]
\label{algo:CKT_randomForests_bad}
\SetAlgoLined
    \KwIn{Initial dataset $\Dc = (X_{i,1}, X_{i,2}, \Z_i)_{i = 1, \dots, n}
        \in \left( \Rb^{2+p}\right)^n$ ;}
    Compute the dataset of pairs $\tilde \Dc$ using Algorithm \ref{algo:creation_dataset_D_tilde} on $\Dc$ \;
    \For{$j \leftarrow 1$ \KwTo $N_{tree}$} {
        Sample a set $\Sc_j \subset \{1, \dots, n(n-1)/2\}$ without replacement \;
        Compute the dataset of pairs $\tilde \Dc_j = (W_k, \tilde Z_k, V_k)_{k \in \Sc_j}$ using observations from $\tilde \Dc$ \;
        Sample a set $\Sc'_j \subset \{ 1, \dots, p' \}$ without replacement \;
        Estimate a classification tree $\Tc_j$ on the dataset $(W_k, (\psi_l(\tilde Z_k))_{l \in \Sc'_j}, V_k)_{k \in \Sc_j}$  \;
    }
    \KwOut{An estimator $\hat \tau_{1,2|\Z = \cdot} := N_{tree}^{-1} \sum_{i=1}^{N_{tree}} \Tc_j(\cdot)$.}
    \caption{Random forests un-adapted for the estimation of the conditional Kendall's tau}
\end{algorithm}

\begin{algorithm}[htb]
\label{algo:CKT_randomForests}
\SetAlgoLined
    \KwIn{Initial dataset $\Dc = (X_{i,1}, X_{i,2}, \Z_i)_{i = 1, \dots, n}
        \in \left( \Rb^{2+p}\right)^n$ ;}
    \For{$j \leftarrow 1$ \KwTo $N_{tree}$} {
        Sample a set $\Sc_j \subset \{1, \dots, n\}$ without replacement \;
        $\Dc_j \leftarrow (X_{i,1}, X_{i,2}, \Z_i)_{i \in \Sc_j}$ \;
        Compute the dataset of pairs $\tilde \Dc_j = (W_k, \tilde Z_k, V_k)_{k \in \Kc_j}$ using Algorithm \ref{algo:creation_dataset_D_tilde} on $\Dc_j$, providing $\Kc_j$ \;
        Sample a set $\Sc'_j \subset \{ 1, \dots, p' \}$ without replacement \;
        Estimate a classification tree $\Tc_j$ on the dataset $(W_k, (\psi_l(\tilde Z_k))_{l \in \Sc'_j}, V_k)_{k \in \Kc_j}$  \;
    }
    \KwOut{An estimator $\hat \tau_{1,2|\Z = \cdot} := N_{tree}^{-1} \sum_{i=1}^{N_{tree}} \Tc_j(\cdot)$.}
\caption{Random forests adapted for the estimation of the conditional Kendall's tau}
\end{algorithm}

\FloatBarrier

\subsection{Nearest neighbors}

\mds

The nearest neighbors provide also a very popular classification algorithm and it can be used directly on the dataset $\tilde \Dc$ (see Algorithm \ref{algo:CKT_nearest_neighbors}). Here, we no longer assume~(\ref{model:cond_tau_Z}) or even~(\ref{model:param_cond_tau}), and we live in a nonparametric framework. A pretty difficult problem is to choose a convenient number of nearest neighbors. As usual in nonparametric statistics, we must find a compromise between variance (tendancy to undersmooth, i.e. to choose a too small $N$) and bias (tendancy to oversmooth, i.e. to choose a too big $N$).
Moreover, in our case, with $n(n-1)/2$ possible pairs, choosing a right value for $N$ can be challenging.
Indeed, in the usual (iid) nearest neighbor framework, the asymptotically optimal $N$ is a power of the sample size. 
Here, this is different because there are three potential sample sizes: $n$, if we consider there are fundamentally $n$ sources of randomness,  $n(n-1)/2$ by considering that the new sample has a cardinality equal to the number of pairs, or even $|\Kc|$ that is random and depends on $h$. Thus, our problem is to choose a ``relevant formula'' for $N$ based on the ``convenient'' sample size.

\begin{algorithm}[htb]
\label{algo:CKT_nearest_neighbors}
\SetAlgoLined
    \vspace{0.1cm}
    \KwIn{A dataset of pairs $\tilde \Dc := (W_k, \tilde \Z_k, V_k)_{k \in \Kc}$}
    \KwIn{A point $\z \in \Zc$, a number $N$ of nearest neighbors and a distance $d$ on $\Rb^{p'}$;}
    $\Kc_\z \leftarrow \arg \min_{E \subset \Kc, |E| = N}
    \Big( \sum_{k \in E} d \big(\psibm(\z), \psibm(\tilde \Z_k) \big) \Big)$ \;
    \vspace{0.1cm}
    \KwOut{An estimator $\hat \tau_{1,2|\Z = \z}^{(N)}
    := \big( \sum_{k \in \Kc_\z} V_k W_k \big) / \sum_{k \in \Kc_\z} V_k$.}
\caption{Estimation of the conditional Kendall's tau $\tau_{1,2|\Z = \z}$ using nearest neighbors.}
\end{algorithm}

In applications, one might not be interested in the value of the conditional Kendall's tau at only one point, but also in the whole function $\z \mapsto \tau_{1,2|\Z = \z}$. The goodness of this estimation is linked to the underlying density $f_\Z$ of $\Z$: the estimation can be made more precise in regions where $f_\Z$ is high, allowing to use a higher number of neighbors with close covariates. At the opposite, in regions where $f_\Z$ is low, a smaller $N$ should be used. Note that, in general, $f_\Z$ is unknown and its estimation may be difficult as well, due to the curse of dimensionality. Therefore, it is highly desirable to build a local number of neighbors $N(\z)$.
Such a local choice $N(\z)$ will help to avoid both under- and over-smoothing in all parts of the space $\Zc$.

\mds

Cross-validation techniques are widely use for the choice of tuning parameters, but might not the best solution here as one would like to find a local choice of $N$. 
This problem has similarities with the classical non-parametric regression, and we propose to use a procedure inspired by Lepski's method for choosing the bandwidth \cite{lepski1997optimal}, once adapted to our setting.
Lepski's method is built on a simple principle: when two non-parametric estimators are close, the best is the smoothest. When two non-parametric estimators are far apart, the best is the least smooth. Let $(\Zc_i)_{i \in \Ic}$ be a partition of $\Zc$. The goal will be to choose the best estimator on each $\Zc_i$, which corresponds to the choice of a local number of nearest neighbors $N_i$.
This procedure is called ``local'' since the diameters of the $\Zc_i$ will be small. For example, if $p=1$ and $\Zc$ is a bounded interval then the $\Zc_i$ can be chosen as small intervals. 
We denote by $\Nc \subset \Nb$ the finite set of possible numbers of neighbors. Following Lepski's approach, we choose $\Nc$ as a geometric progression, i.e. $\Nc = \{ \lfloor a_1 \times a_2^i \rfloor, \, i=1, \dots, i_{max} \}$ for some constants $a_1, a_2 > 0$, where $\lfloor x \rfloor$ denotes the integer part of a real $x$.

\mds

To measure how far the estimators are from each other, we introduce a distance $d_i$, $i\in \Ic$. As our estimators of conditional Kendall's tau are bounded (between $-1$ and $1$) and measurable, several choices are possible. In applications, we will use
\begin{equation}
    d_i(f,g) = \bigg(\frac{1}{j_{max}} \sum_{j = 1}^{j_{max}} \Big[(f(\z_{i,j}) - g(\z_{i,j})) / M \Big]^2 \bigg)^{1/2}, \z_{i,1} , \dots, \z_{i,j_{max}} \in \Zc_i,
    \label{def:d_i_distance_estim_CKT}
\end{equation}
where $M$ is a normalization factor independent of $i$ and the subsets $\z_{i,1} , \dots, \z_{i,j_{max}}$ are arbitrarily chosen in $\Zc_i$, $i=1,\ldots,i_{max}$. We will use
$M = (\max - \min) \{ \hat \tau_{1,2|\Z=\z}^{(N)} , N \in \Nc, \z \in \Zc \}$.
Indeed, in the classical nonparametric regression $Y=f(X)+\varepsilon$, with unknown $f$ to estimate, $M$ should be replaced by the standard deviation of the noise $\varepsilon$. In our case, we can define a (pseudo-)noise
$\xi_{\z, N} := \hat \tau_{1,2|\Z = \z}^{(N)} - \tau_{1,2|\Z = \z}$, but it is unknown in practice and its distribution is complicated.
Therefore $M$ serves as a proxy of the amplitude of the variations in the estimated conditional Kendall's tau. This normalization by $M$ ensures a kind of adaptativity of the estimation.

\mds

\begin{algorithm}[htb]
\label{algo:CKT_knn_Lepski}
\SetAlgoLined
    \vspace{0.1cm}
    \KwIn{A set $\Nc \subset \Nb$ of possible numbers of nearest neighbors and the corresponding estimates $\hat \tau_{1,2|\Z= \cdot}^{(N)}$ given by Algorithm \ref{algo:CKT_nearest_neighbors}, for all $N \in \Nc$;}
    \KwIn{A partition $(\Zc_i)_{i \in \Ic}$ of $\Zc$ and a distance $d_i$ on a space of bounded measurable real functions defined on $\Zc_i$, for every $i \in \Ic$;}
    \ForEach{$i \in \Ic$} {
        \vspace{-0.3cm}
        \begin{align}
            &\hspace{-1.4cm} S_i \leftarrow \Big\{ N \in \Nc: d_i \Big( \hat \tau_{1,2|\Z= \cdot}^{(N)} \, , \, \hat \tau_{1,2|\Z= \cdot}^{(N')} \Big)
            \leq A \sqrt{ (1/N') \log(max(\Nc) / N')},
            \forall N' \in \Nc \cap[1, N] \Big\} \; ; 
            \label{eq:def_set_S_i} \\
            &\hspace{-1.4cm} N_i \leftarrow \max(S_i) ;  \nonumber
        \end{align}
        \vspace{-0.7cm}
    }
    \KwOut{An estimator $\z \mapsto \hat \tau_{1,2|\Z = \z}
    := \sum_{i \in \Ic} \1\{\z \in \Zc_i \} \hat \tau_{1,2|\Z=\z}^{(N_i)}$.}
\caption{Lepski's method for a local choice of the number of nearest neighbors, and the corresponding estimator of the conditional Kendall's tau.}
\end{algorithm}

We have observed that sensitivity to $\Nc$ is not too large, if it is chosen in a reasonable way, for example between $5$ or $10$ possibilities. When $\Z$ is univariate, a simple partition $(\Zc_i)_{i \in \Ic}$ can be given by the deciles of $\Z$. We choose $A = 1$ for simplicity since we believe there is no procedure for choosing it. A statistician which would like to play with the smoothness of the result is free to adapt $A$, using an expert knowledge of the situation. Finally, the $\z_{i,j}$ can be chosen as quantiles of $\Zc$, or as a regular grid on each $\Zc_i$.

\subsection{Neural networks}
\label{subsection:neural_network}

\mds

Nowadays, neural networks have become very popular with a wide range of applications. 
In classification problems, a neural network can be seen as an estimator depending on some parameters, but in a very flexible and complex way. For every input $\z$, it yields the probability of belonging to any class.
In our framework, we will train a network on the dataset of pairs $\tilde \Dc$. It is well-known that most neural networks do not induce convex programs, and the outputs therefore depend on some initial parameter values. One strategy is to independently train networks with different starting parameter values, that may be randomly chosen, for example.

\mds

This method of using independent estimators (conditionally on the initial sample $\Dc$) and then aggregating them is related to the random forest approach of the previous section and the discussion therein. 
Therefore, the same techniques are relevant and we have noticed an improvement in performance by using an adapted version of Algorithm \ref{algo:CKT_randomForests}.
More precisely, we fix a number of neural networks. For each neural network, we sample without replacement a part of the initial dataset from which the corresponding dataset of pairs is constructed and used as a training set. In order to improve stability, we aggregate the predictions of the different neural networks by using the median as the final output. There is a trade-off between computation time and accuracy: a larger number of networks should improve the accuracy while taking obviously a longer time to be trained.
The precise choice of the best architecture of the network is a complicated task, which is left for future research. As we are looking for functions $\z \mapsto \tau_{1,2|\Z=\z}$ which are smooth almost everywhere and easy to interpret in applications, we choose a simple architecture with $10$ neural networks, each having one hidden layer of $3$ neurons. Besides, bigger networks seem to deteriorate the performance of this estimator, see Section~\ref{subsection:simu_choice_number_neurons}.

\begin{algorithm}[htb]
\label{algo:CKT_nnet}
\SetAlgoLined
    \KwIn{Initial dataset $\Dc = (X_{i,1}, X_{i,2}, \Z_i)_{i = 1, \dots, n}
        \in \left( \Rb^{2+p}\right)^n$ ;}
    \For{$j \leftarrow 1$ \KwTo $N_{nnet}$} {
        Sample a set $\Sc_j \subset \{1, \dots, n\}$ without replacement \;
        $\Dc_j \leftarrow (X_{i,1}, X_{i,2}, \Z_i)_{i \in \Sc_j}$ \;
        Compute the dataset of pairs $\tilde \Dc_j = (W_k, \tilde Z_k, V_k)_{k \in \Kc_j}$ using Algorithm \ref{algo:creation_dataset_D_tilde} on $\Dc_j$, providing $\Kc_j$ \;
        Estimate a neural net $\mathfrak{N}_j$ on the dataset $(W_k, \psibm(\tilde Z_k), V_k)_{k \in \Kc_j}$  \;
    }
    \KwOut{An estimator $\hat \tau_{1,2|\Z = \cdot} := 
    Median \{ \mathfrak{N}_j (\cdot), j = 1, \dots, N_{nnet} \}$.}
\caption{Neural networks with median bagging, adapted for the estimation of the conditional Kendall's tau}
\end{algorithm}

\subsection{Lack of independence and its influence on the proposed algorithms}
\label{subsection:influence_independence}

\mds

The machine learning methods that are adapted in this section were all designed for i.i.d. data. But it is easy to see that observations in the dataset of pairs $\tilde \Dc$ will not be independent.
Indeed, assume that observations in the original dataset $(X_{i,1}, X_{i,2}, \Z_i)_{i = 1, \dots, n}$ are i.i.d., to simplify.
The pair $(i=1,j=2)$ and the pair $(i=1,j=3)$ both involve the first observation $(X_{1,1}, X_{1,2}, \Z_1)$, and therefore are not independent.
This is a theoretical problem, but numerical results in Section~\ref{section:simulation} show that this is not a problem in practice.

\mds

As far as the logit and probit are concerned, it was proved in the previous Section~\ref{section:reg_approach} that they are related to a family of estimators that can use $\tilde \Dc$ ``as is'', and that nonetheless yield consistent and asymptotically normal estimates, if the specification is good. It is likely that the other methods presented here enjoy similar properties and are also largely unaffected by dependence between pairs.
Note that, if all observations in $\Dc$ are identically distributed, then observations in $\tilde \Dc$ are identically distributed as well. 
This is in favor of our methods.

\mds

Concerning the dependence inside $\tilde \Dc$, we will show that it is not too strong. For example, the pairs $(1,2)$ and $(1,3)$ are not independent, but the pairs $(1,2)$ and $(3,4)$ are indeed independent. This means that there is still ``a large proportion of'' independence left in $\tilde \Dc$. Formally, if two distinct pairs are randomly chosen in $\tilde \Dc$, the probability of them being independent is high. Indeed, there are 
$N_{tot} :=  n(n-1)(n(n-1)-2) / 8$ couples of distinct pairs. Beside, the number $N_{ind}$ of couples of pairs which are independent is 
$N_{ind} := n (n-1)(n-2)(n-3)/8.$ The factor $1/8$ appears in both $N_{tot}$ and $N_{ind}$ since we can always switch the two observations in the first pair, in the second pair, and switch the two pairs (every $4$-tuple is counted $2^3 = 8$ times).
It is easy to see that $N_{ind} / N_{tot} = 1 - O(1/n)$ as $n \to \infty$. 

\mds

This can be interpreted in the following way: as $n \to \infty$, pairs are ``almost all'' independent from each other. In other words, the dependence between two pairs become negligible with averages. That is the reason why the following machine learning methods will be performing well if the original dataset $\Dc$ is large enough. If the original dataset $\Dc$ is not iid itself, for example coming from a time series, we conjecture that such methods will work in a similar way as long as the dependence is not too strong, for example if the data-generating process satisfies some usual assumptions, see Remark~\ref{rem:temporal_dependence}.

\mds

Whenever bootstrap, subsetting, resampling or cross-validation is done on these classification-based estimators, we advise to perform them on the original dataset $\Dc$ rather than on the dataset of pairs $\tilde \Dc$, as we do in Sections~\ref{subsection:logit_probit_classifiers}, \ref{subsection:trees_random_forests} and \ref{subsection:neural_network}. This seems to yield a good improvement in performance. An example is given by the difference between Algorithms~\ref{algo:CKT_randomForests_bad} and~\ref{algo:CKT_randomForests}. This can be simply summed up as \textit{``do the resampling on the original dataset $\Dc$, not on the transformed dataset $\tilde \Dc$''}. Nevertheless, a complete study and justification of this general principle is beyond the scope of this paper and is left for future work.

\section{Simulation study}
\label{section:simulation}

In this section, we have studied the relative performances of our estimators by simulation. 
For a given model and a given method of estimation, we sample 100 different experiments, and estimate the model for each sample. We fix the sample size as $n=3000$.
We remark that for a given dimension $p>0$ of $\Z$ and a given support $\Zc$ of $\Z$, we have different ``blocks'' of the model which can be chosen in an independent way:
\begin{enumerate}[(i)]
    \item the law $\PP_\Z$ of $\Z$ ;
    \item the function $\z \in \Zc \mapsto \tau_{1,2|\Z=\z}$ ;
    \item the (conditional) copula family $(C_\tau)_{\tau \in (0,1)}$ or $(C_\tau)_{\tau \in (-1,1)}$ of $(X_1,X_2)|\Z=\z$, indexed by its conditional Kendall's tau. For example the Gaussian, Student, Clayton, Gumbel, etc, copula families. Such a family can also depend on $\Z$: for example, think of a Student copula with varying degrees of freedom~;
    \item the conditional margins $X_1|\Z$ and $X_2|\Z$ ;
    \item the choice of the functions $\psi_i$, for $i=1,\ldots,p'$ ;
    \item the choice of the estimator $\hat \tau_{1,2|\Z = \cdot}$.
\end{enumerate}

Our so-called ``reference setting'' will be defined as $p=1$, $\Zc=[0,1]$ and
(i) $\PP_Z = \Uc_{[0;1]}$ ;
(ii) $\tau_{1,2|\Z=\z} =  3z(1-z)$ ;
(iii) $(C_\tau)_{\tau \in (0,1)} = GaussianCopula$ ;
(iv) $\PP_{X_1|Z=z} = \PP_{X_2|Z=z} = \Nc(z,1)$.
For each tested model, the performance of the estimator will be evaluated by the mean integrated $\ell_2$ error. With obvious notations, it will be estimated as
\begin{equation}
    Err := \EE \left[\int_{\Zc} (\hat \tau_{1,2|\Z=\z} - \tau_{1,2|\Z=\z})^2 d\z \right]
    \approx \frac{1}{N_{simu}  N_{points}} \sum_{i=1}^{N_{simu}} \sum_{j=1}^{N_{points}} (\hat \tau_{1,2|\Z=\z^{(j)}}^{(i)} - \tau_{1,2|\Z=\z^{(j)}})^2,
    \label{eq:def:criteria_l2}
\end{equation}
where $N_{simu}, N_{points}$ are positive integers, $\z^{(1)}, \dots, \z^{(N_{points})}$ are fixed points in $\Zc$, and $\hat \tau_{1,2|\Z=\z^{(j)}}^{(i)}$ is the estimated conditional Kendall's tau at point $\z^{(j)}$ trained on data from the $i$-th simulation. We choose $N_{simu}:=100$ experiments, and in this reference setting, the integral is discretized with $N_{points}:=100$ points equispaced on the segment $[0.01 , 0.99]$ to avoid boundary numerical problems.

\mds

In the following simulations, ``Logit'' and ``Probit'' refer to Algorithm \ref{algo:CKT_logit_probit}.
``Tree'' refers to the application of the method \texttt{tree()} of package \texttt{tree} by Ripley \cite{ripley2018tree} on the dataset $\tilde \Dc$ produced by Algorithm \ref{algo:creation_dataset_D_tilde}.
``Random forests'' refers to Algorithm \ref{algo:CKT_randomForests}.
``Nearest neighbors'' refers to the adapted version using Algorithm \ref{algo:CKT_nearest_neighbors}, once aggregated using Algorithm \ref{algo:CKT_knn_Lepski}.
Finally ``Neural networks'' refers to Algorithm \ref{algo:CKT_nnet}. Such specifications are now part of our ``reference setting''.

\subsection{Choice of $\{ \psi_i \}, i=1, \dots, p'$.}
\label{subsection:choice_psibm}

\mds

We consider six different choices of $\psibm$, that are

\begin{enumerate}
    \item No transformation, i.e. $\psi_1^{(1)}(z)=z$ ;
    
    \item Polynomials of degree lower than 4: $\psi_{i}^{(2)}(z) = 2^{-i+1} (z-0.5)^{i-1}$ for $i=1,\dots, 5$ ;
    
    \item Polynomials of degree lower than 10: $\psi_{i}^{(3)}(z) = 2^{-i+1} (z-0.5)^{i-1}$ for $i=1,\dots, 11$ ;
    
    \item Fourier basis of order 2 with an intercept: $\psi_1^{(4)}(z) = 1$, 
    $\psi_{2i}^{(4)}(z) = \cos(2\pi i z)$ and
    $\psi_{2i+1}^{(4)}(z) = \sin(2\pi i z)$ for $i=1,2$ ;
    
    \item Fourier basis of order 5 with an intercept: $\psi_1^{(5)}(z) = 1$, 
    $\psi_{2i}^{(5)}(z) = \cos(2\pi i z)$ and
    $\psi_{2i+1}^{(5)}(z) = \sin(2\pi i z)$ for $i=1,\dots, 5$ ;
    
    \item Concatenation of $\psibm^{(2)}$ and $\psibm^{(4)}$, which will be denoted by $\psibm^{(6)}$.
\end{enumerate}

For each of the choices of $\psibm$ above, and each estimator, we compute the criteria (\ref{eq:def:criteria_l2}).
The results are displayed in the following Table \ref{tab:result_simu_psibm}.

\begin{table}[htb]
    \centering
    
    \begin{tabular}{c|cccccc}
        Chosen $\psibm$ & $\,$ Logit $\,$ & $\,$ Probit $\,$
        & $\,$ Tree $\,$ & $\,$ Random forests $\,$ & $\,$
        Nearest neighbors $\,$ & $\,$ Neural network $\,$ \\
        \hline 
        $\psibm^{(1)}$ & 48.1 & 48.1 &  7.5 & 4.89 & 2.26 & 0.561  \\ 
        $\psibm^{(2)}$ & 0.721 & 0.554 & 4.28 & 3.28 & 2.26 & 1.32  \\ 
        $\psibm^{(3)}$ & 0.663 & 0.528 & 4.13 & 3.41 & 2.23 & 1.73  \\ 
        $\psibm^{(4)}$ & 1.41 & 1.45 & 4.73 & 14.2 & 2.72 & 1.74  \\ 
        $\psibm^{(5)}$ & 1.05 & 1.06 & 4.76 & 10.3 & 2.79 & 2.67  \\ 
        $\psibm^{(6)}$ & 0.456 & 0.434 & 4.57 & 3.15 & 2.64 & 3.87
    \end{tabular}
    
    \caption{Error criteria (\ref{eq:def:criteria_l2}) for each choice of $\psibm$ and each method, multiplied by $1000$.}
    
    \label{tab:result_simu_psibm}
\end{table}

With the choice of $\psibm^{(6)}$, Logit and Probit methods provide the best results. This good performance deteriorates with other choices of $\psibm$, especially when the model is misspecified. Neural networks provide the best results with $\psibm^{(1)}$, and their performance declines when further transformations of $\z$ are introduced in $\psibm$. Nearest neighbors have nearly the best behavior with $\psibm^{(1)}$, and it does not seem that other transformations can significantly increase its performance. On the contrary, for trees and random forests, it seems that bigger families $\psibm$ can yield improvements over $\psibm^{(1)}$.


\mds

\begin{figure}[htbp]
    {\centering
    \includegraphics[width = 0.8\textwidth]{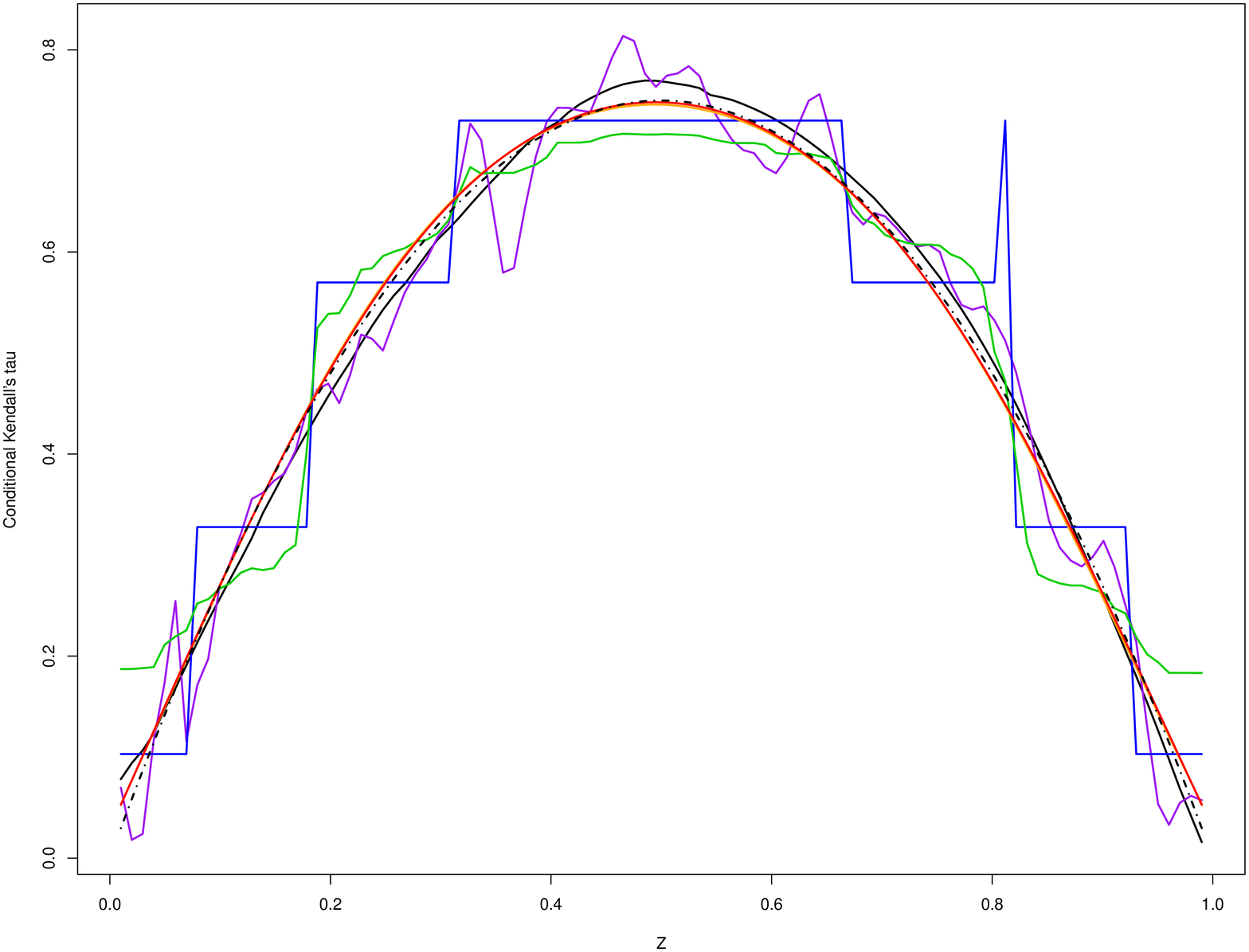}
    \caption{An example of the estimation of the conditional Kendall's tau using different estimation methods (see Table \ref{tab:correspondence_colors} below). The black dash-dotted curve is the true conditional Kendall's tau that has been used in the simulation experiment.}
    
    \label{fig:classification_comparison}
    
    \vspace{1cm}
    
    \renewcommand{\arraystretch}{1.5}
    
    \begin{tabular}{c|cccccc}
        Method & $\,$ Logit $\,$ & $\,$ Probit $\,$
        & $\,$ Tree $\,$ & $\,$ Random $\,$ & $\,$
        Nearest $\,$ & $\,$ Neural $\,$ \\[-0.2cm]
        & $\,$ $\,$ & $\,$ $\,$
        & $\,$ $\,$ & $\,$ forests $\,$ & $\,$
        neighbors $\,$ & $\,$ network $\,$ \\
        \hline
        Algorithm & Algorithm \ref{algo:CKT_logit_probit}
        & Algorithm \ref{algo:CKT_logit_probit}
        & \texttt{tree()} of \cite{ripley2018tree}
        & Algorithm \ref{algo:CKT_randomForests}
        & Algorithms \ref{algo:CKT_nearest_neighbors}-\ref{algo:CKT_knn_Lepski}
        & Algorithm \ref{algo:CKT_nnet} \\
        \hline
        Color & orange & red & blue & green & purple & black \\
    \end{tabular}
    
    \captionof{table}{Summary of available estimation methods for the estimation of the conditional Kendall's tau and corresponding algorithm and curve color.}
    
    \label{tab:correspondence_colors}
    }
    
    \mds
    
    For each estimator, we precise in the second line of the above table the algorithm used to compute it, and in the third line the color of the corresponding curve on Figures~\ref{fig:classification_comparison} to~\ref{fig:CKT_X1_X4_Z2_P4}.
    For example, the estimator ``Probit'' is computed using Algorithm~\ref{algo:CKT_logit_probit} and correspond to the red curves.

\end{figure}

From now on, we will choose $\psibm^{(6)}$ for the methods \textit{logit}, \textit{probit}, \textit{tree} and \textit{random forests}. Indeed, for these methods, the choice of $\psibm$ yield nearly the lowest error criteria and present the advantage of being diverse, which will help to combine the performances of both polynomials and oscillating functions.
On the contrary, for the methods \textit{nearest neighbors} and \textit{neural networks}, we choose $\psibm^{(1)}$ as adding new functions does not seem to increase the performance of both of these methods.
Figure \ref{fig:classification_comparison} displays a comparison of the different methods on a typical simulated sample.

\subsection{Comparing different copulas families}

\mds

Here, we keep the reference setting and we change only part (iii), i.e. the functional form of the conditional copula.
The results are displayed in Table \ref{tab:result_simu_copulaFamily}.
We observe that such choice of a parametric copula families has nearly no effect on the performance of the estimators.
Nonetheless, with the Student copula - either with fixed or variable degrees of freedom - most estimators have slightly worse performances than with other copulas. It can be explained by the fact that this copula allows asymptotic dependence, i.e. a strong tail association.

\begin{table}[htbp]
    \centering
    
    \begin{tabular}{c|cccccc}
        Copula family & $\,$ Logit $\,$ & $\,$ Probit $\,$
        & $\,$ Tree $\,$ & $\,$ Random $\,$ & $\,$
        Nearest $\,$ & $\,$ Neural $\,$ \\
        & $\,$ $\,$ & $\,$ $\,$
        & $\,$ $\,$ & $\,$ forests $\,$ & $\,$
        neighbors $\,$ & $\,$ network $\,$ \\
        \hline 
        Gaussian & 0.456 & 0.434 & 4.57 & 3.15 & 2.26 & 0.561  \\ 
        Student $4$ df & 0.549 & 0.515 & 4.54 & 3.28 & 2.87 & 0.753  \\ 
        Student $(2+1/z)$ df & 0.531 & 0.518 & 4.66 & 3.23 & 2.82 & 0.805  \\ 
        Clayton & 0.498 & 0.472 & 4.52 & 3.36 & 2.67 & 0.742  \\ 
        Gumbel & 0.45 & 0.431 & 4.56 & 3.23 & 2.66 & 0.775  \\ 
        Frank & 0.448 & 0.42 &  4.5 & 3.28 & 2.13 & 0.615 
    \end{tabular}
    
    \caption{Error criteria (\ref{eq:def:criteria_l2}) for each copula family and each method, multiplied by $1000$.}
    
    \label{tab:result_simu_copulaFamily}
\end{table}

\subsection{Comparing different conditional margins}

In this subsection, we stil start from the reference setting and we change only part (iv), i.e. the functional form of the conditional margins $(X_1|Z)$ and $(X_2|Z)$.
We consider the following alternatives:
\begin{enumerate}
    \item $\PP_{X_1|Z=z} = \PP_{X_2|Z=z} = \Nc(z,1)$ (as in the reference case).
    
    \item $\PP_{X_1|Z=z} = \Nc( \cos(10 \pi z), 1) \; ; \;
    \PP_{X_2|Z=z} = \Nc(z, 1)$. The idea is to make $X_1$ oscillate fast so that the algorithms will have difficulties to localize concordant and discordant pairs ;
    
    \item $\PP_{X_1|Z=z} = Exp(|z|) \; ; \;
    \PP_{X_2|Z=z} = \Uc_{[z, z+1]}$. 
    This choice allows to see how estimation is affected by changes in the conditional support of $(X_1, X_2)$ given $\Z=\z$ ;
    
    \item $\PP_{X_1|Z=z} = \Nc(0, z^2) \; ; \;
    \PP_{X_2|Z=z} = \Uc_{[0, |z|]}$. 
    Then, we will see how estimation is affected by changes in the conditional variance of $(X_1, X_2)$ given $\Z=\z$.
\end{enumerate}

\begin{table}[htb]
    \centering
    
    \begin{tabular}{c|cccccc}
        Setting & $\,$ Logit $\,$ & $\,$ Probit $\,$
        & $\,$ Tree $\,$ & $\,$ Random forests $\,$ & $\,$
        Nearest neighbors $\,$ & $\,$ Neural network $\,$ \\
        \hline 
        1.  & 0.456 & 0.434 & 4.57 & 3.15 & 2.26 & 0.561  \\ 
        2.  & 0.809 & 0.818 & 4.65 & 3.72 & 2.65 & 0.838  \\ 
        3.  & 1.15 & 1.12 & 5.29 & 4.21 & 3.57 & 1.32  \\ 
        4.  & 0.493 & 0.471 & 4.43 & 3.44 & 2.54 & 0.662 
    \end{tabular}
    
    \caption{Error criteria (\ref{eq:def:criteria_l2}) for each choice of conditional margins and each method, multiplied by $1000$.}
    
    \label{tab:result_simu_condMargins}
\end{table}

In a similar way as in the previous section, the results of these experiments, as displayed in Table \ref{tab:result_simu_condMargins} show that changes in terms of conditional marginal distributions generally have a mild impact on the overall performance of the estimators. Moreover, such changes have no effect on the ranking between estimators: \textit{logit} and \textit{probit} methods are always the best, followed by the \textit{neural networks}. \textit{Nearest neighbors}, and \textit{random forests} are behind in this order. The estimator \textit{Tree} shows the lowest performance, but note that it also has the lowest computation time.

\subsection{Comparing different forms for the conditional Kendall's tau}

In this part, we keep the reference setting, but we change only part (ii), i.e. the functional form of the conditional Kendall's tau itself.
We consider the following choices:
\begin{enumerate}
    \item $f_1(z) := 0.9 - 0.8 \, \1\{z \geq 0.5\}$,
    \item $f_2(z) := 3z(1-z)$,
    \item $f_3(z) := 0.5 + 0.4 \sin(4 \pi z)$,
    \item $f_4(z) := 0.1 + 1.6 z \1\{z < 0.5\} + 1.6(z-0.5)\1\{u \geq 0.5\}$.
\end{enumerate}
The  results are presented in Table \ref{tab:result_simu_form_condKendallTau}.
If the estimated model is close to be well-specified, the best methods are parametric, i.e. the \textit{logit} and \textit{probit} regressions.
In all the other cases, \textit{neural networks} seem to perform very well. There is a compromise between minimization of the error, and minimization of the computation time. We refer to Table~\ref{tab:result_simu_influence_n} for a quantitative comparison of the performance of the methods in terms of computation time, as a function of the sample size $n$.

\begin{table}[htb]
    \centering
    
    \begin{tabular}{c|cccccc}
        Setting & $\,$ Logit $\,$ & $\,$ Probit $\,$
        & $\,$ Tree $\,$ & $\,$ Random forests $\,$ & $\,$
        Nearest neighbors $\,$ & $\,$ Neural network $\,$ \\
        \hline 
        $f_1$  & 11.2 & 11.6 & 4.12 & 4.03 & 3.89 & 1.48  \\ 
        $f_2$  & 0.456 & 0.434 & 4.57 & 3.15 & 2.26 & 0.561  \\ 
        $f_3$  & 3.77 & 3.22 & 5.95 & 4.76 & 2.35 & 2.17  \\ 
        $f_4$  & 12.8 & 12.8 & 16.8 &   10 & 3.71 & 1.97
    \end{tabular}
    \caption{Error criteria (\ref{eq:def:criteria_l2}) for different Kendall's tau models and each estimation method, multiplied by $1000$.}
    
    \label{tab:result_simu_form_condKendallTau}
\end{table}

\subsection{Higher dimensional settings}

In the previous sections, we had chosen a univariate vector $\Z$, i.e. $p=1$. 
Since this may sound a bit restrictive, we would like to obtain some finite-sample results in dimension $p=2$. Note that the latter dimension cannot be too high because of the curse of dimensionality linked with the necessary kernel smoothing (done in Algorithm \ref{algo:creation_dataset_D_tilde} when creating the dataset of pairs). We also choose a simple dictionary $\psibm$ of functions, which will consists in the two projections on the coordinates of $\Z$.
The performance of the estimators is still be assessed by the approximate error criteria (\ref{eq:def:criteria_l2}). The corresponding $\z^{(j)}$ are  chosen as a grid of $400$ points equispaced on the square $[0.01 , 0.99]^2$.

\mds

In this framework, we first choose block (iii) of the model : the conditional copula of $X_1$ and $X_2$ given $\Z$ will be Gaussian, and block (iv) : $\PP_{X_1|\Z=\z} = \PP_{X_2|\Z=\z} = \Nc(z_1,1)$.
We will try different combinations for the remaining blocks (i) and (ii), described as follows:

\begin{enumerate}[(1)]
    \item $Z_1 \sim \Nc(0,1),$ $Z_2 \sim \Uc_{[-1,1]},$ and the copula of $(Z_1, Z_2)$ is Gaussian with a Kendall's tau equal to $0.5$. 
    Moreover, $\tau_{1,2|\Z=\z} = z_2  \tanh(z_1)$. This model is interesting because the function $\z \mapsto \tau_{1,2|\Z=\z}$ will be far away from a linear function of $\psibm(\z)$, and the machine learning techniques should work better than logistic/probit regressions.
    
    \item we keep the same model as previously, but by setting $g(\tau_{1,2|\Z=\z}) = z_1 + z_2$, using the $g$ of Example \ref{example:logit} so that we recover the parametric setting of Section \ref{section:reg_approach}.
    
    \item $Z_1 \sim Exp(1)$, $Z_2 \sim \Nc(0,1)$ and both variables are independent. Set $\tau_{1,2|\Z=\z} = \exp(-z_1 |z_2|)$. Again, we have a misspecified nonlinear model that is far away from logit/probit models.
\end{enumerate}
The results are given in Table~\ref{tab:result_simu_dimZ_2_psibm1}. With the exception of the well-specified setting (2), the logit model performs worse than non-parametric methods. In all these settings, neural networks show better performances than all other methods, followed by nearest neighbors and tree-based methods. Finally, parametric methods are the worst, especially under misspecification of the model.

    
    

\begin{table}[htb]
    \centering
    
    \begin{tabular}{c|cccccc}
        Setting & $\,$ Logit $\,$ & $\,$ Probit $\,$
        & $\,$ Tree $\,$ & $\,$ Random forests $\,$ & $\,$
        Nearest neighbors $\,$ & $\,$ Neural network $\,$ \\
        \hline 
        (1) & 35.5 & 35.5 & 9.63 & 11.7 & 6.72 & 2.21  \\ 
        (2) & 0.433 & 0.681 & 10.9 & 5.85 & 4.33 & 0.848  \\ 
        (3) & 17.8 & 17.2 & 5.72 & 9.79 & 1.84 & 1.36  \\ 
    \end{tabular}
    
    \caption{Error criteria (\ref{eq:def:criteria_l2}) for each setting with $2$-dimensional $\Z$ random vectors and each method, multiplied by $1000$.}
    
    \label{tab:result_simu_dimZ_2_psibm1}
\end{table}

\subsection{Choice of the number of neurons in the one-dimensional reference setting}
\label{subsection:simu_choice_number_neurons}

\mds

We consider networks with different number of neurons, and study their performance, both statistically and computationally. The results are displayed in the following Table~\ref{tab:result_simu_nnet_neurons}.
We observe that increasing the number of neurons only seems to deteriorate the performance of the method. 

\begin{table}[htb]
    \centering
    
    \begin{tabular}{c|cccc}
        Number of neurons & 3 & 5 & 10 & 30 \\
        \hline
        Criteria & 0.561 & 0.808 & 1.47 & 1.45  \\ 
        Time (s) &  234 &  429 &  607 & 5.29e+03  \\ 
    \end{tabular}
    
    \caption{Error criteria (\ref{eq:def:criteria_l2}) multiplied by $1000$, and average computation time in seconds for each architecture of the neural networks.}
    
    \label{tab:result_simu_nnet_neurons}
\end{table}

\subsection{Influence of the sample size $n$}

\mds

In our one-dimensional reference setting, we fix all the parameters except $n$. For a grid of values of $n$, we evaluate the performance of our estimators.

\begin{table}[htb]
    \centering
    \renewcommand{\arraystretch}{1.2}
    \begin{tabular}{cc|cccccc}
        & & $\,$ Logit $\,$ & $\,$ Probit $\,$
        & $\,$ Tree $\,$ & $\,$ Random $\,$ & $\,$
        Nearest $\,$ & $\,$ Neural $\,$ \\
        && $\,$ $\,$ & $\,$ $\,$
        & $\,$ $\,$ & $\,$ forests $\,$ & $\,$
        neighbors $\,$ & $\,$ network $\,$ \\
        \hline
        \multirow{2}{*}{$n=1000$} & Criteria & 1.58 & 1.52 & 5.85 & 4.45 & 4.01 & 2.01  \\ 
        & Time (s) & 59.6 &  156 & 0.215 & 8.11 & 5.04 & 30.6  \\ 
        \hline
        \multirow{2}{*}{$n=2000$} & Criteria & 0.666 & 0.64 &  4.9 & 3.39 & 2.95 & 1.79  \\ 
        & Time (s) &  192 &  489 & 0.99 & 35.9 & 17.1 & 85.3  \\ 
        \hline
        \multirow{2}{*}{$n=3000$} & Criteria & 0.456 & 0.434 & 4.57 & 3.15 & 2.26 & 0.561 \\
        & Time (s) & 414 & 1010 & 2.37 & 87 & 36.9 & 234  \\ 
        \hline
        \multirow{2}{*}{$n=5000$} & Criteria & 0.275 & 0.253 & 3.77 & 3.05 & 1.69 & 0.791  \\ 
        & Time (s) &  957 &  2420 & 6.37 &  218 &  111 &  461  \\ 
        \hline
        \multirow{2}{*}{$n=8000$} & Criteria & 0.22 & 0.204 &  3.6 & 3.39 & 1.27 & 0.225  \\ 
        & Time (s) &  2178 &  5480 & 15.2 &  499 &  290 &  1268  \\ 
    \end{tabular}
    
    \caption{Error criteria (\ref{eq:def:criteria_l2}) multiplied by $1000$ and computation time in seconds for each method and each choice of $n$.}
    
    \label{tab:result_simu_influence_n}
\end{table}

We observe that, for most methods, the computation time increases and the error criteria decreases when the sample size increases.
We note that at most, the number of pairs is $n(n-1)$, and therefore, the computation time should increase as $O(n^2)$, which is coherent with the results of Table~\ref{tab:result_simu_influence_n}.
The relative order of the performances does not seem to change with the sample size $n$ : the same methods are the best ones with small or large $n$.
Note that we have not tried to find an ``optimal'' fine-tuning of the parameters, for each method and each choice of $n$. 
Indeed, to find optimal choices of tuning parameters is not an easy task (on a theoretical and practical sense), 
and more precise studies are left for future research.

\subsection{Influence of the lack of independence}

\mds

In Section~\ref{subsection:influence_independence}, we detail theoretical considerations about the lack of independence in the dataset $\tilde \Dc$ and some consequences. The following simulation experiment complements this analysis with the following empirical results.

\mds

Indeed, when using Algorithm~\ref{algo:creation_dataset_D_tilde}, we note that pairs of observations are not independent, and therefore, the elements of the dataset of pairs $\tilde \Dc$ are not independent from each other. This should degrade the performance of our methods, compared to a situation where all elements would be independent. We now consider such a situation, in order to compare the performance of the methods in both cases. Note that the cardinality of $\tilde \Dc$ is $n(n-1)/2$.
Therefore, we will compare the two following settings:
\begin{enumerate}
    \item Reference situation: fix $n=3000$, simulate $n$ independent copies
    $\Dc_n := (X_{i,1}, X_{i,2},\Z_i)_{i=1, \dots, n}$, construct the dataset of pairs $\tilde \Dc_n$ using Algorithm~\ref{algo:creation_dataset_D_tilde}. Use the estimators on the training set $\tilde \Dc_n$.
    \item Independent situation: fix $n=3000$, simulate $n(n-1) \simeq 9,000,000$ independent copies $\Dc_{n(n-1)} := (X_{i,1}, X_{i,2},\Z_i)_{i=1, \dots, n(n-1)}$. Create the dataset of consecutive pairs $\bar\Dc_{n(n-1)}$ on this sample using Algorithm~\ref{algo:creation_dataset_D_bar}. This means that we use only consecutive pairs, i.e. (1,2), (3,4), (5,6), and so on. 
    Use the estimators on the training set $\bar\Dc_{n(n-1)}$.
\end{enumerate}

\begin{algorithm}[htb]
\label{algo:creation_dataset_D_bar}
\SetAlgoLined
    \vspace{0.1cm}
    \KwIn{Initial dataset $\Dc = (X_{i,1}, X_{i,2}, \Z_i)_{i = 1, \dots, n}
    \in \left( \Rb^{2+p}\right)^n$ ;}
    \For{$k \leftarrow 1$ \KwTo $\lfloor n \rfloor /2$} {
        $i,j \leftarrow 2k-1, 2k$ \;
        $\tilde \Z_k \leftarrow (\Z_i + \Z_j) / 2$ \;
        $W_k \leftarrow W_{(i,j)}$ as defined in Equation (\ref{eq:def_W_ij}) \;
        $V_k \leftarrow K_h(\Z_i - \Z_j)$ \;
    }
    Define $\Kc := \{k: V_k > 0\}$ \;
    \KwOut{A dataset of pairs $\bar \Dc := (W_k, \tilde \Z_k, V_k)_{k \in \Kc}
    \in \big( \{-1, 1\} \times \Rb^p \times \Rb_+
    \big)^{\lfloor n \rfloor /2}$.}
\caption{Algorithm for creating the dataset of consecutive pairs from the initial dataset.}
\end{algorithm}

Note that, by construction, the cardinalities of $\bar\Dc_{n(n-1)}$ and $\tilde \Dc_n$ are the same, i.e. both have exactly $n(n-1)/2$ pairs. 
This is the reason why we chose to simulate $n(n-1)$ points in the independent situation, so that these two numbers of pairs can match.
Note that the elements in $\bar \Dc$ are independent from each other by construction while some elements in $\tilde \Dc$ may not be independent from each other, in general.
We can now compare the performances of the estimators trained on $\bar\Dc_{n(n-1)}$ and on $\tilde \Dc_n$ using the criteria (\ref{eq:def:criteria_l2}).
Some results are given in Table~\ref{tab:result_comparion_independence}.
Note that the simulation of the each $(X_{i,1}, X_{i,2},\Z_i)$ is still made under the previous one-dimensional ``reference setting''.

\begin{table}[htb]
    \centering
    \renewcommand{\arraystretch}{1.2}
    \begin{tabular}{c|cccccc}
        & $\,$ Logit $\,$ & $\,$ Probit $\,$
        & $\,$ Tree $\,$ & $\,$ Random $\,$ & $\,$
        Nearest $\,$ & $\,$ Neural $\,$ \\
        & $\,$ $\,$ & $\,$ $\,$
        & $\,$ $\,$ & $\,$ forests $\,$ & $\,$
        neighbors $\,$ & $\,$ network $\,$ \\
        \hline
        Independent & 0.127 & 0.114 & 3.02 & 2.52 & 0.12 & 0.0363 \\
        Not independent & 0.456 & 0.434 & 4.57 & 3.15 & 2.26 & 0.561  \\ 
    \end{tabular}
    
    \caption{Error criteria (\ref{eq:def:criteria_l2}) multiplied by $1000$ for each method and each situation.
    ``Independent'' means the independent situation with $\bar \Dc_{n(n-1)}$, and ``Not independent'' means the reference situation with $\tilde \Dc_n$.}
    
    \label{tab:result_comparion_independence}
\end{table}

As expected, all estimators show a better performance in the independent situation. Nonetheless, the independent situation has been simulated using $n(n-1) \simeq 9,000,000$ points whereas the reference situation uses only $n=3,000$ points. Even if the numbers of pairs in both experiments are the same, the sample size of the dataset was much larger in the independent situation. This means that there is more information available, and explains also why the independent situation has a better performance: it just uses more data. Such a huge sample may not be available in practice though.
    
\mds

Nevertheless, the original procedure costs $O(n^2)$, which can be large for very large values of $n$. In this case, it is always to possible to restrict oneself to consecutive pairs, with a cost of only $O(n)$. Such a procedure is possible if the dataset is very large and Algorithm~\ref{algo:creation_dataset_D_bar} can be seen as an alternative to Algorithm~\ref{algo:creation_dataset_D_tilde} where only consecutive pairs are used. This would lower the computation cost, at the expense of precision.

\mds

\section{Applications to financial data}
\label{section:applications_financial}

In this section, we study the changes in the conditional dependence between the daily returns of MSCI stock indices during two periods: the European debt crisis (from 18 March 2009 to 26 August 2012) and the after-crisis period (26 August 2012 to 2 March 2018).
We will consider the following couples : (Germany, France), (Germany, Denmark), (Germany, Greece), respectively denoted by $(X_1, X_2), (X_1, X_3), (X_1, X_4)$.
We will separately consider two choices of conditioning variables $\Z$: 
\begin{itemize}
    \item a proxy variable for the intraday volatility $\sigma := (High - Low) / Close$, where $High$ denotes the maximum daily value of the Eurostoxx index, $Low$ denotes its minimum and $Close$ is the index value at the end of the corresponding trading day. 
    
    \item a proxy of so-called ``implied volatility moves''  $\Delta\sigma^I$. It will record the daily variations of the EuroStoxx 50 
    Volatility Index, whose quotes are available at \url{https://www.stoxx.com/index-details?symbol=V2TX}: $\Delta\sigma^I_{i} := V2TX(i) - V2TX(i-1)$ for each trading day $i$. 
    The EuroStoxx 50 Volatility Index $V2TX$ measures the levels of future volatility, as anticipated by the market through option prices.
\end{itemize}

\mds

Note that, for a given couple, the levels of the estimated conditional Kendall's tau are different (in general) for different conditioning variables. Indeed, the unconditional Kendall's tau $\tau_{1,2}$, the average conditional Kendall's tau with respect to $\sigma$, which is $\EE_{\sigma} \big[ \tau_{1,2|\sigma} \big]$ and the average conditional Kendall's tau with respect to $\Delta\sigma^I$, which is $\EE_{\Delta\sigma^I} \big[ \tau_{1,2|\Delta\sigma^I} \big]$ have no reason to be equal.

\mds

Both conditioning variables $\sigma$ and $\Delta\sigma^I$ are of dimension $1$. For each method and each conditioning variable, we will use the ``best'' choice of $\psibm$ as determined from the simulations in Section \ref{subsection:choice_psibm}, that is $\psibm^{(6)}$ for the methods 
\textit{logit}, \textit{probit}, \textit{tree} and \textit{random forests} and $\psibm^{(1)}$ for the methods \textit{Nearest neighbors} and  \textit{neural networks}. On the following figures, the matching between colors and corresponding estimators still follows Table~\ref{tab:correspondence_colors}.

\begin{rem}
    \label{rem:temporal_dependence}
    It is well-known that sequences of asset returns are not iid. In particular, their volatilities are time-dependent, as in GARCH-type or stochastic volatility models. Moreover, the tail behavior of their distributions are significantly varying, due to some periods of market stress. Several families of models (switching regime models, jumps, etc) have tried to capture such stylized facts. We conjecture that such temporal dependencies will not affect too much our results. Indeed, dependence will be mitigated by considering all possible couples of random vectors, independently of their dates. It is easy to go one step beyond, for instance by keeping only the 
    couples of returns indexed by $i$ and $j$ when $|i-j|>m$, for some ``reasonably chosen'' threshold $m$ ($m=20$, e.g.). 
    In every case, it is highly likely that our inference procedures are still consistent and asymptotically normal, for most types of dependence between successive observations (mixing processes, weak dependence, $m$-dependence, mixingales, etc), even if the asymptotic variances are different from ours.   
\end{rem}

\begin{figure}[p]
    \begin{minipage}[c]{0.48\textwidth}
        \includegraphics[width=1.05\textwidth, height = 6cm]{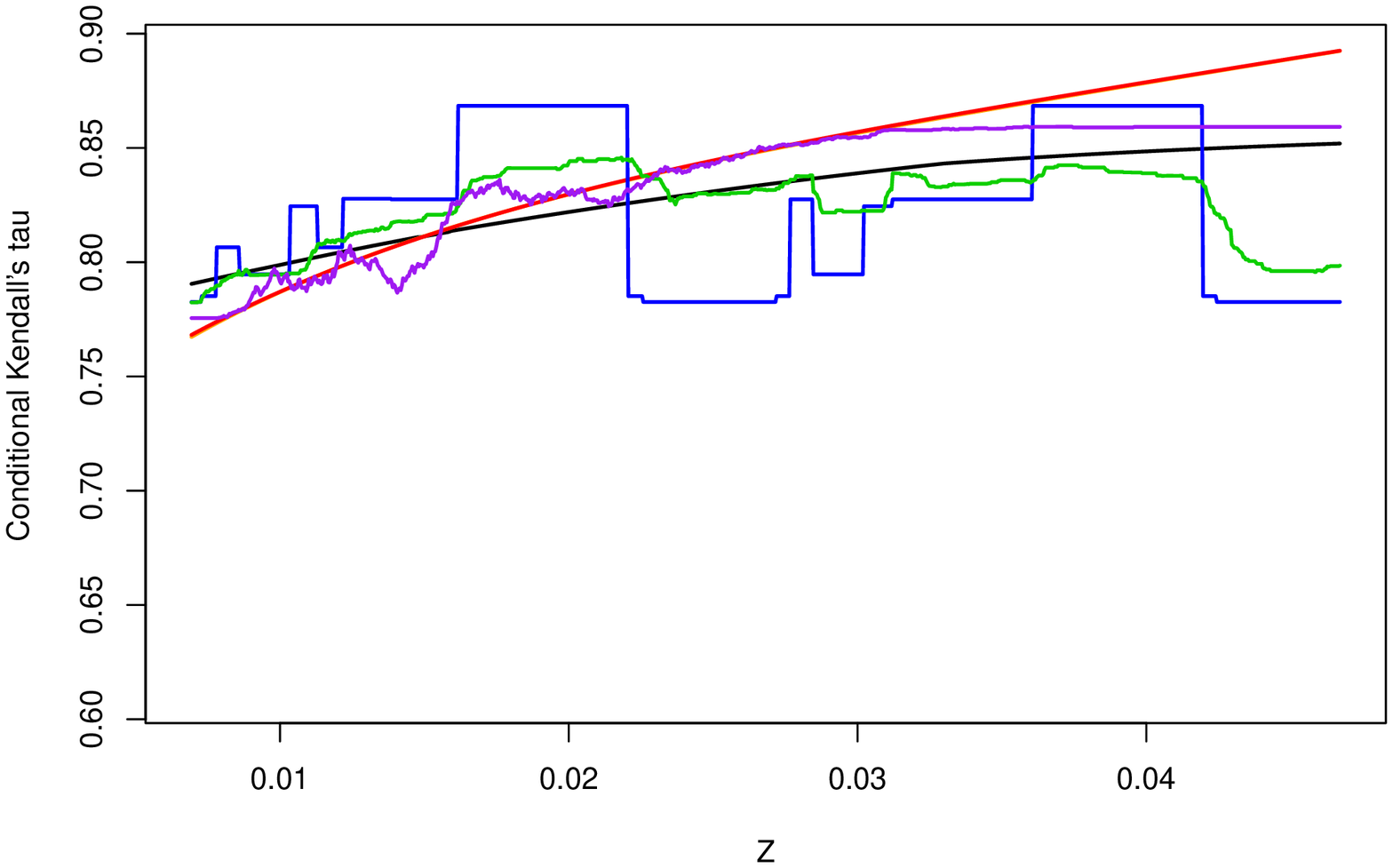}
        \caption{Conditional Kendall's tau between $(X_1, X_2)$ given $\sigma$ during the European debt crisis}
        \label{fig:CKT_X1_X2_Z1_P3}
    \end{minipage}
    \hfill
    \begin{minipage}[c]{0.48\textwidth}
        \includegraphics[width=1.05\textwidth, height = 6cm]{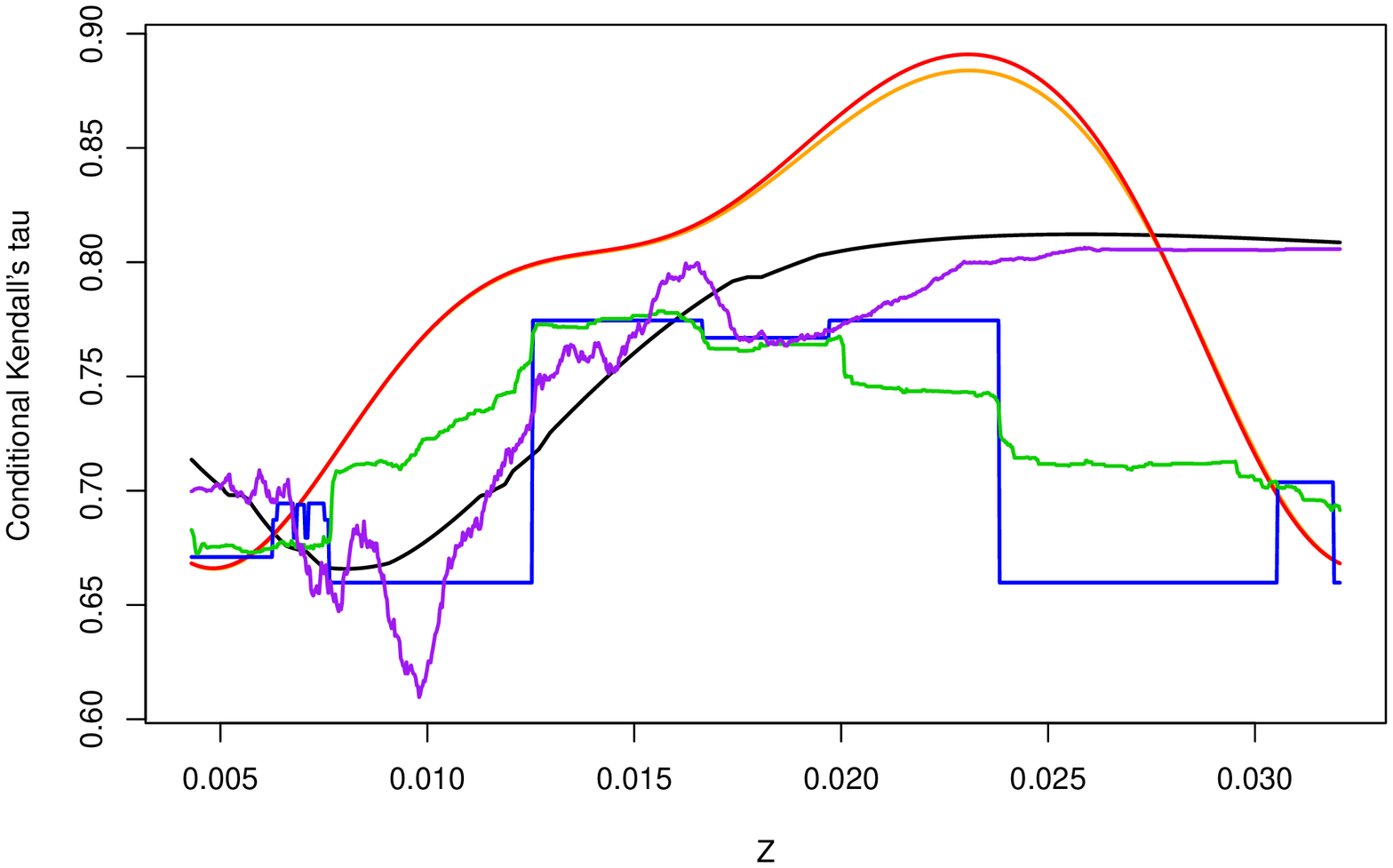}
        \caption{Conditional Kendall's tau between $(X_1, X_2)$ given $\sigma$ during the After-crisis period}
        \label{fig:CKT_X1_X2_Z1_P4}
    \end{minipage}
    \begin{minipage}[c]{0.48\textwidth}
        \includegraphics[width=1.05\textwidth, height = 6cm]{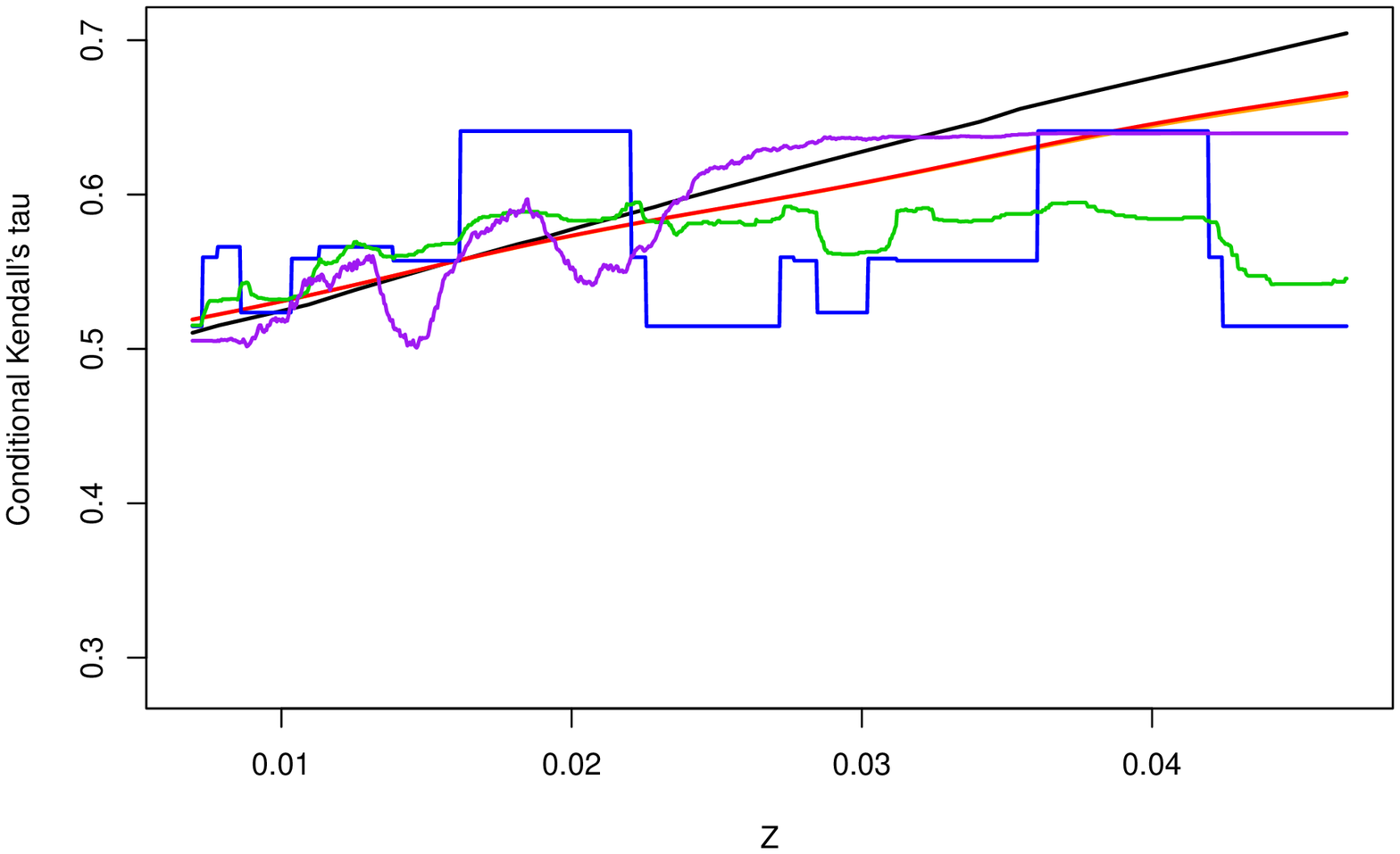}
        \caption{Conditional Kendall's tau between $(X_1, X_3)$  given $\sigma$ during the European debt crisis}
        \label{fig:CKT_X1_X3_Z1_P3}
    \end{minipage}
    \hfill
    \begin{minipage}[c]{0.48\textwidth}
        \includegraphics[width=1.05\textwidth, height = 6cm]{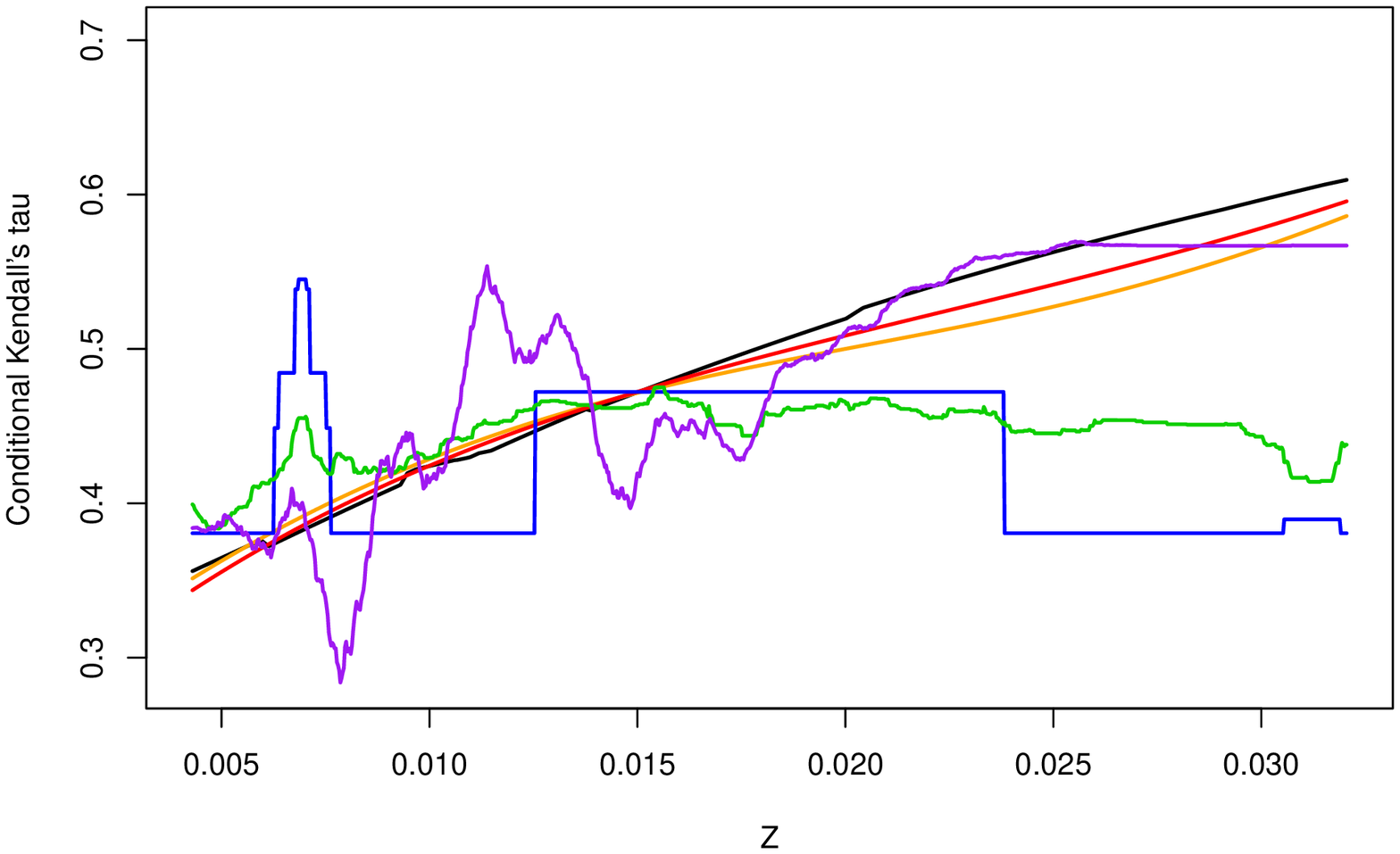}
        \caption{Conditional Kendall's tau between $(X_1, X_3)$ given $\sigma$ during the After-crisis period}
        \label{fig:CKT_X1_X3_Z1_P4}
    \end{minipage}
    \begin{minipage}[c]{0.48\textwidth}
        \includegraphics[width=1.05\textwidth, height = 6cm]{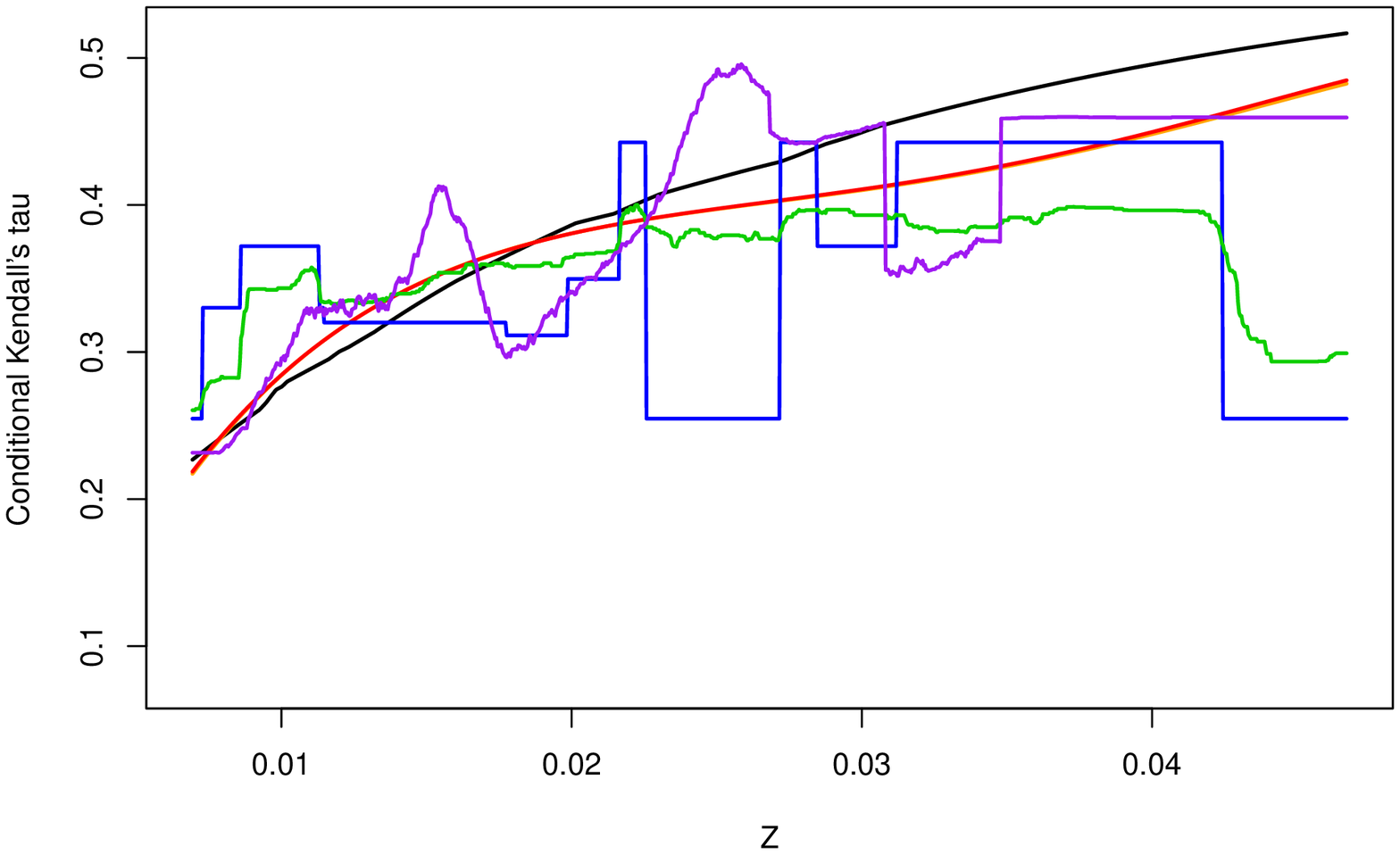}
        \caption{Conditional Kendall's tau between $(X_1, X_4)$  given $\sigma$ during the European debt crisis}
        \label{fig:CKT_X1_X4_Z1_P3}
    \end{minipage}
    \hfill
    \begin{minipage}[c]{0.48\textwidth}
        \includegraphics[width=1.05\textwidth, height = 6cm]{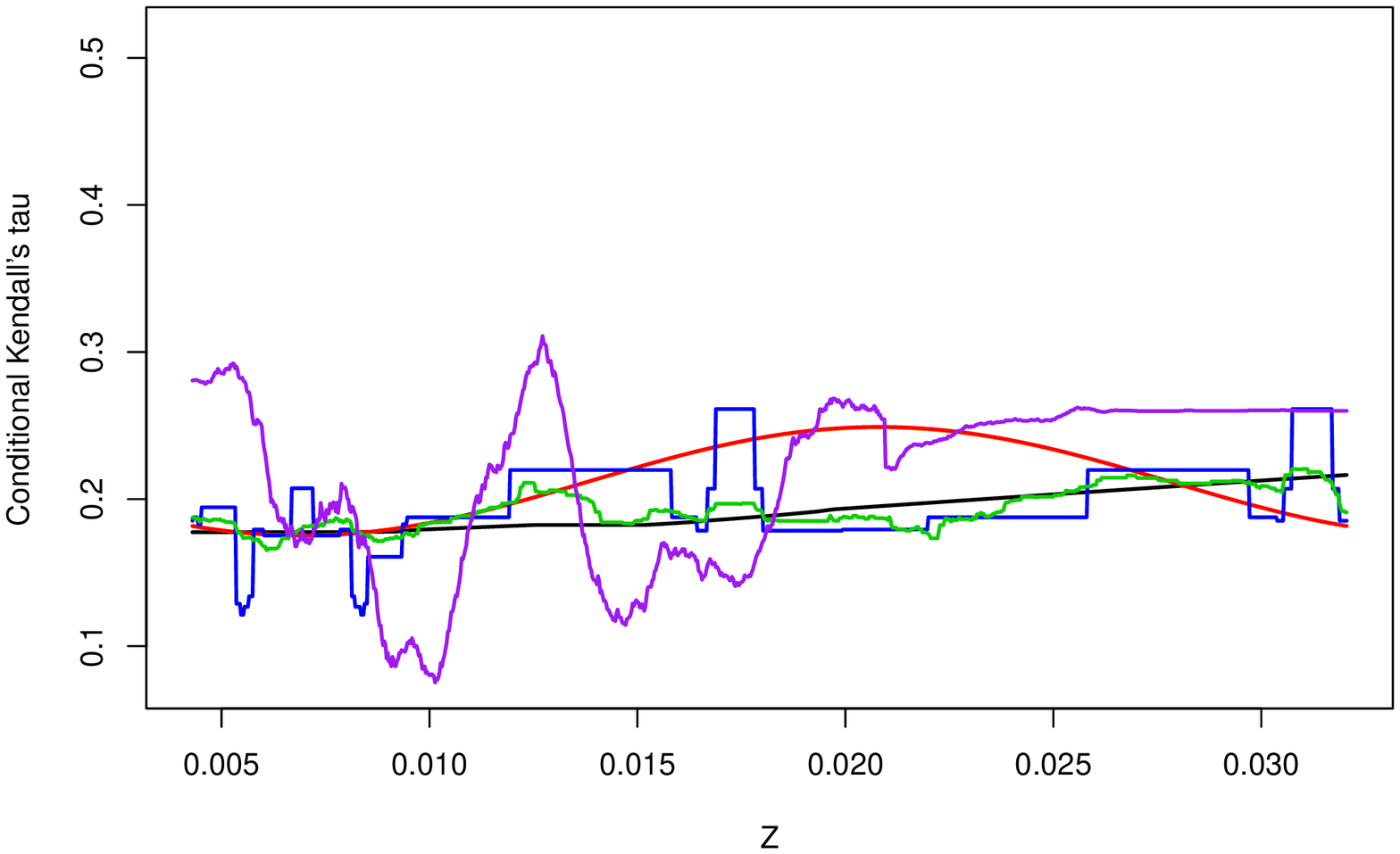}
        \caption{Conditional Kendall's tau between $(X_1, X_4)$ given $\sigma$ during the After-crisis period}
        \label{fig:CKT_X1_X4_Z1_P4}
    \end{minipage}
\end{figure}

\subsection{Conditional dependence with respect to the Eurostoxx's volatility proxy $\sigma$}

We will first consider the conditioning events given by $\sigma$, the proxy variable for the market intraday volatility. The results are displayed on Figures \ref{fig:CKT_X1_X2_Z1_P3} to \ref{fig:CKT_X1_X4_Z1_P4}.
Intuitively, dependence should tend to increase with market volatility: when ``bad news'' are announced, they are sources of stress for most dealers, especially inside the Eurozone that brings together economically connected countries. This phenomenon should be particularly sensitive during the European debt crisis, because a lot of such ``bad news'' were related to the Eurozone itself (economic/financial news of public debts of several European countries). Let us see whether this is the case. 

\mds

On most figures, the estimated conditional Kendall's tau seems to exhibit some kind of concavity. 
The behavior of these functions can be roughly broken down into two main regimes: 
\begin{enumerate}
    \item The ``moderate'' volatility regime (also called the ``normal regime'') in the sense that the volatility stay mild, say in the lower half of its range. In this normal regime, the conditional Kendall's tau is an increasing function of the volatility.
    This is coherent with most empirical research where it is shown that dependence increases with volatility.
    \item The high volatility regime: this is a ``stressed regime'' where $\sigma$ lies in the upper half of its range. In this less frequent regime, the influence of the European volatility $\sigma$ on the conditional Kendall's tau appears to be less clear: the estimators become more ``fluctuating'' and more different from each other, as a consequence of the small number of observations in most stressed regimes.
\end{enumerate}

During the European debt crisis (see Figures \ref{fig:CKT_X1_X2_Z1_P3}, \ref{fig:CKT_X1_X3_Z1_P3} and \ref{fig:CKT_X1_X4_Z1_P3}), the three couples seem to exhibit the same shape of conditional dependence with respect to $\sigma$, even if their average levels are different. These similarities can be a little bit surprising considering that the economic situations of the corresponding countries are different. It can be conjectured that the heterogeneity in the ``mean'' levels of conditional dependence is sufficient to reflect this diversity of situations. In this perspective, the increasing pattern of conditional dependence w.r.t. the ``volatility'' would be a pure characteristic of that period, regardless of the chosen pair of European countries. Indeed, we have observed this pattern for most couples of European countries in the Eurozone.
An explanation might be that investors were focusing on the same international news, for example, about the future of the Eurozone, and they were therefore reacting in a similar way, irrespective of the country.

\mds

For each couple of countries, the conditional Kendall's tau is nearly always lower during the After-crisis period than during the European debt crisis. Apparently, in the after-crisis period, factors and events that are specific to each country attract more attention from investors than during the crisis, which results in lower dependence.
In this context, the shapes of conditional dependence are not the same anymore for different couples.
In particular, the conditional Kendall's tau between German and French returns show a significant increase during the low volatility regime and a decrease during the high volatility regime: see Figure \ref{fig:CKT_X1_X2_Z1_P4}. Between German and Danish returns, the conditional dependence is also increasing during the low volatility regime, but in the high volatility, the conditional Kendall's tau seems to be rather constant, even increasing according to the nearest neighbors and the neural networks estimators.
Concerning Figure \ref{fig:CKT_X1_X4_Z1_P4}, we do not seem any clear tendency. It is likely that $\sigma$ has almost no impact on the conditional dependence between German and Greek stock index returns.

\FloatBarrier

\begin{figure}[htbp]
    \begin{minipage}[c]{0.48\textwidth}
        \includegraphics[width=1.05\textwidth, height = 6cm]{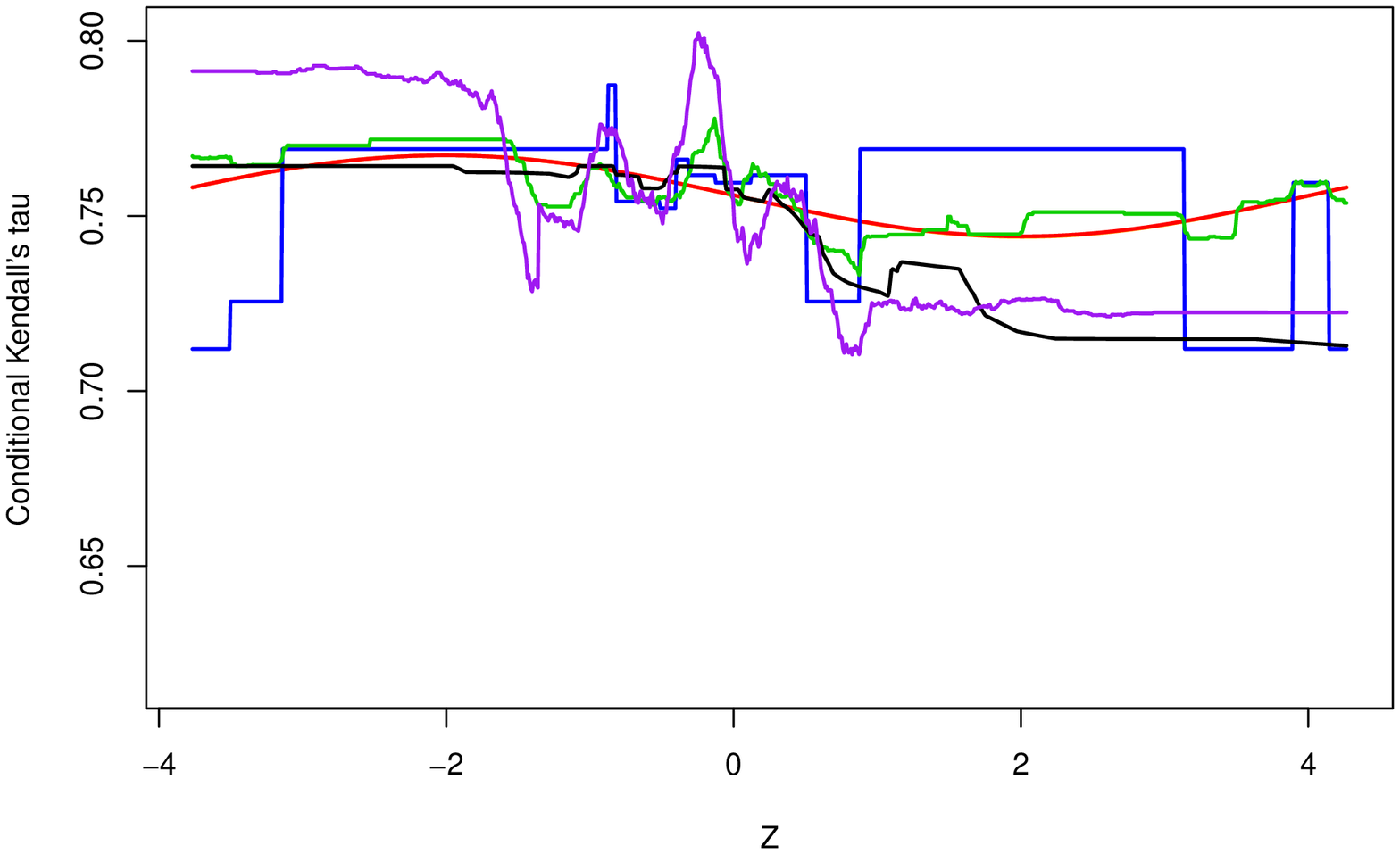}
        \caption{Conditional Kendall's tau between $(X_1, X_2)$ given $\Delta\sigma^I$ during the European debt crisis}
        \label{fig:CKT_X1_X2_Z2_P3}
    \end{minipage}
    \hfill
    \begin{minipage}[c]{0.48\textwidth}
        \includegraphics[width=1.05\textwidth, height = 6cm]{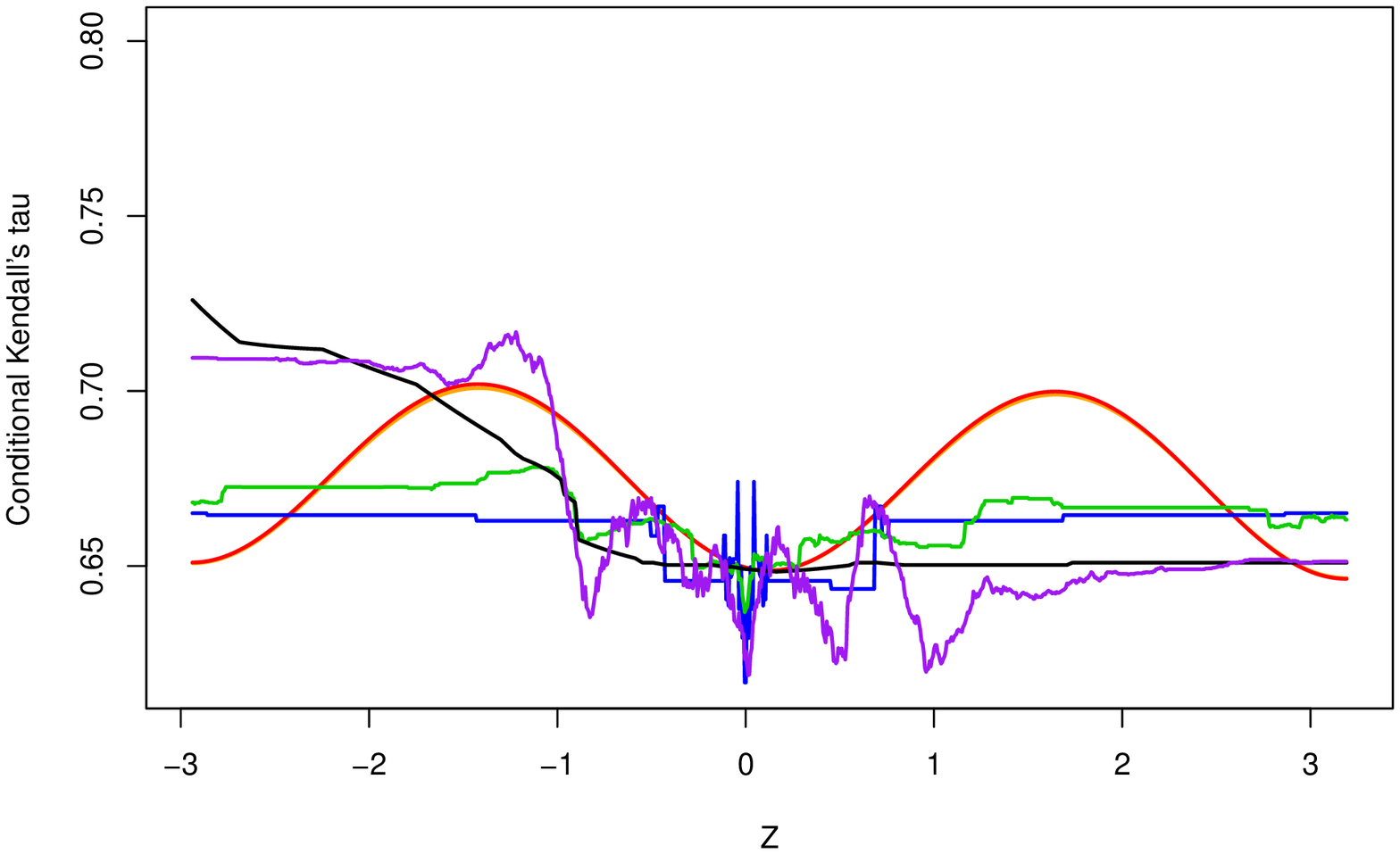}
        \caption{Conditional Kendall's tau between $(X_1, X_2)$ given $\Delta\sigma^I$ during the After-crisis period}
        \label{fig:CKT_X1_X2_Z2_P4}
    \end{minipage}
    \begin{minipage}[c]{0.48\textwidth}
        \includegraphics[width=1.05\textwidth, height = 6cm]{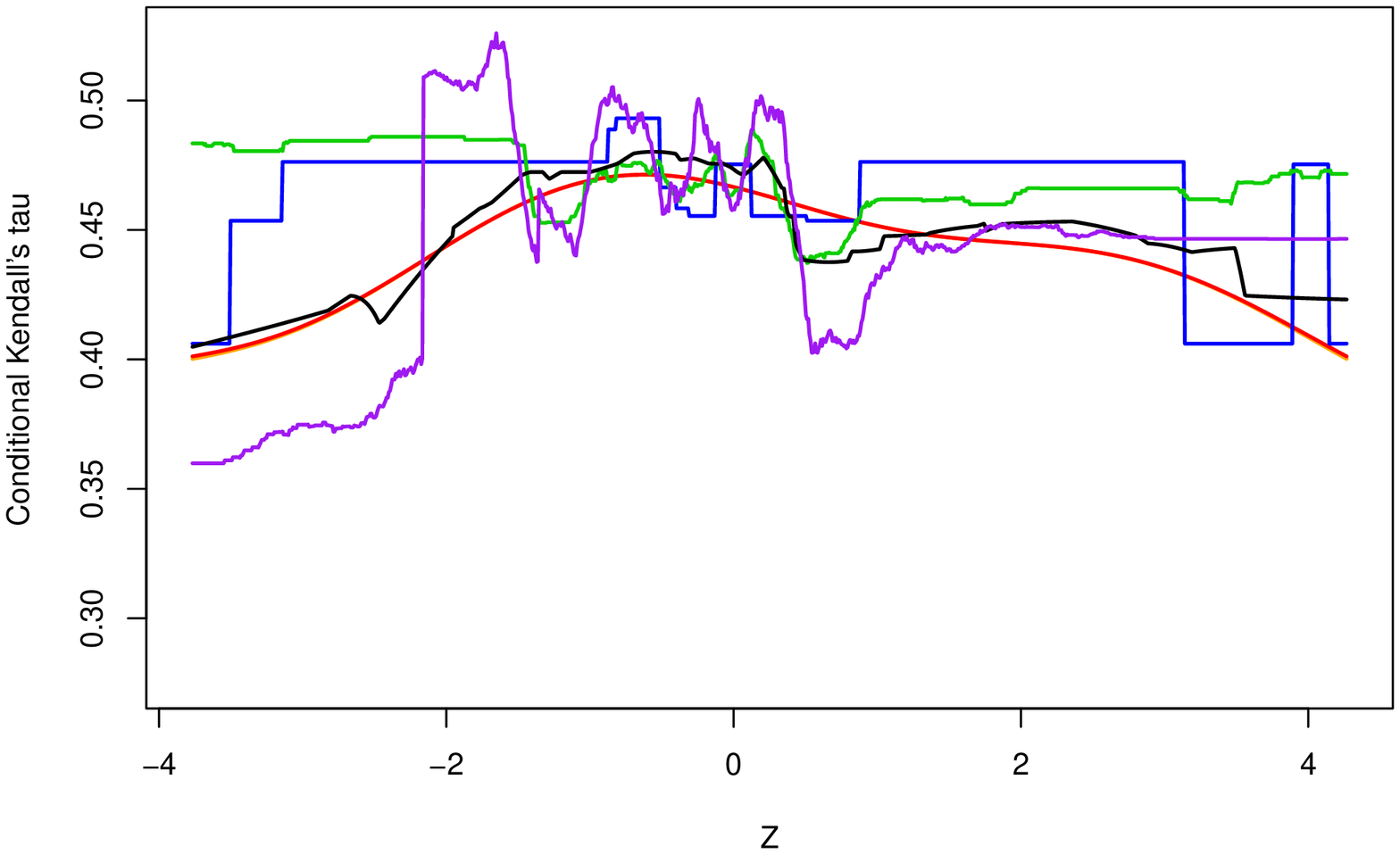}
        \caption{Conditional Kendall's tau between $(X_1, X_3)$ given $\Delta\sigma^I$ during the European debt crisis}
        \label{fig:CKT_X1_X3_Z2_P3}
    \end{minipage}
    \hfill
    \begin{minipage}[c]{0.48\textwidth}
        \includegraphics[width=1.05\textwidth, height = 6cm]{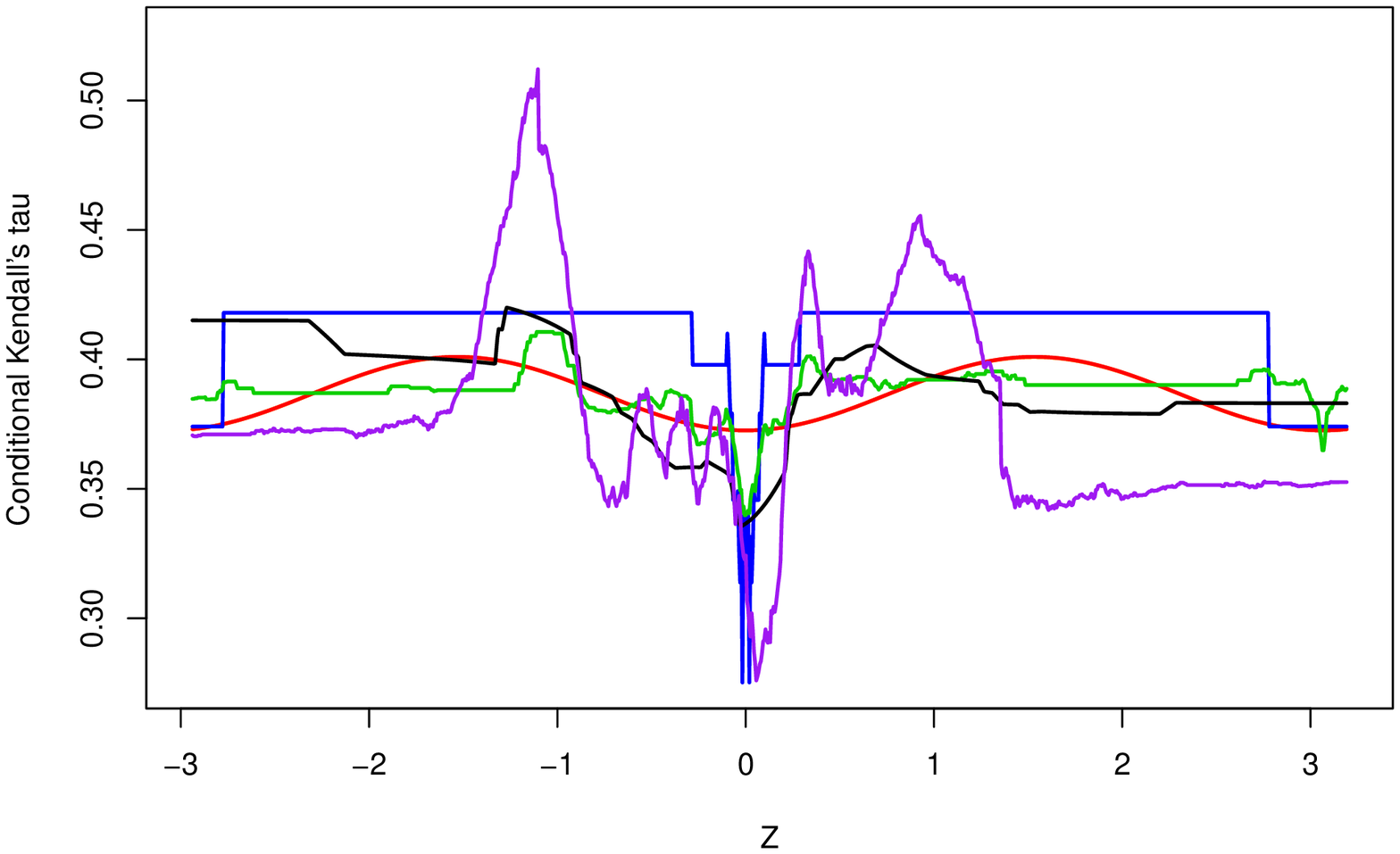}
        \caption{Conditional Kendall's tau between $(X_1, X_3)$ given $ \Delta\sigma^I$ during the After-crisis period}
        \label{fig:CKT_X1_X3_Z2_P4}
    \end{minipage}
    \begin{minipage}[c]{0.48\textwidth}
        \includegraphics[width=1.05\textwidth, height = 6cm]{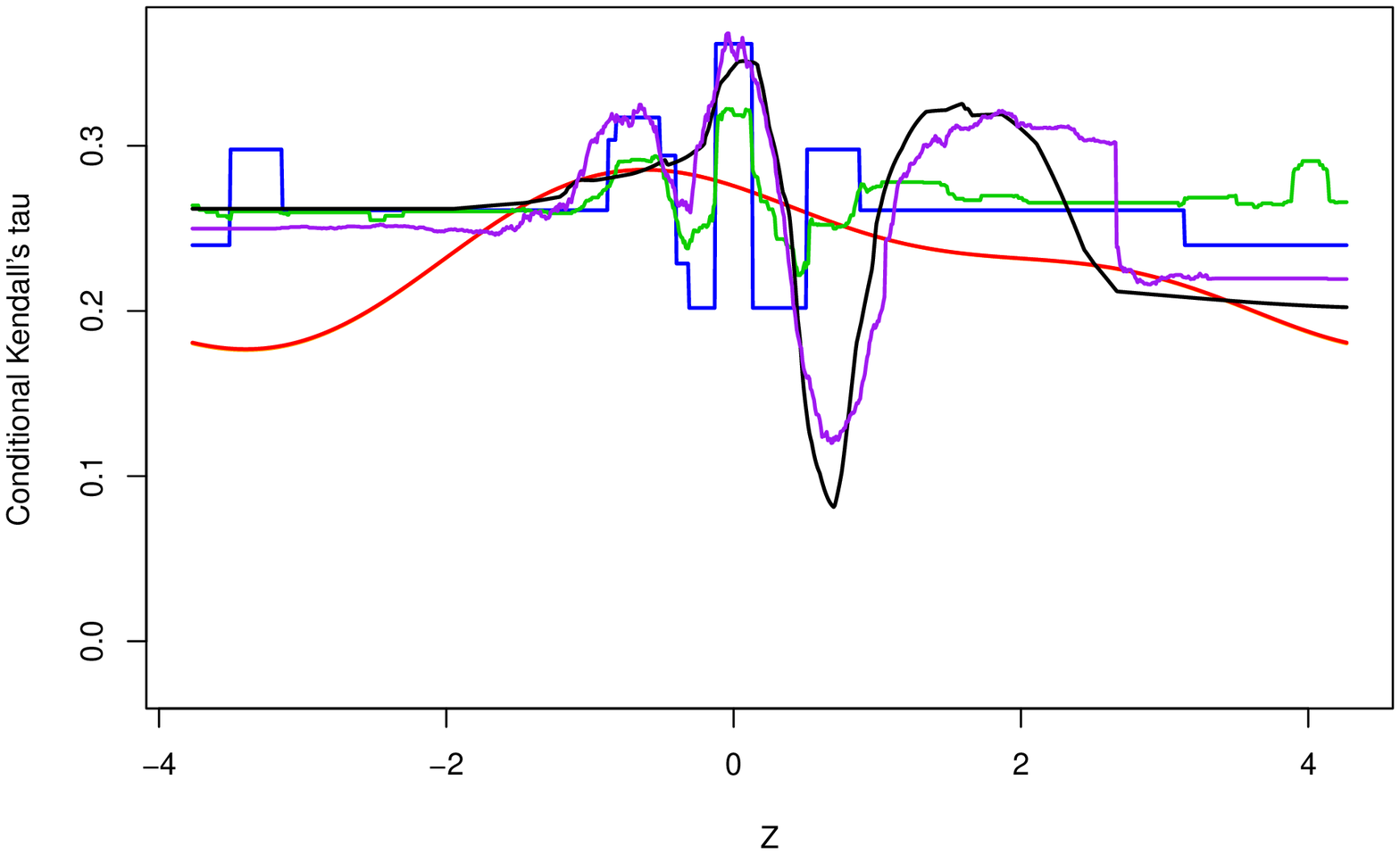}
        \caption{Conditional Kendall's tau between $(X_1, X_4)$ given $\Delta\sigma^I$ during the European debt crisis}
        \label{fig:CKT_X1_X4_Z2_P3}
    \end{minipage}
    \hfill
    \begin{minipage}[c]{0.48\textwidth}
        \includegraphics[width=1.05\textwidth, height = 6cm]{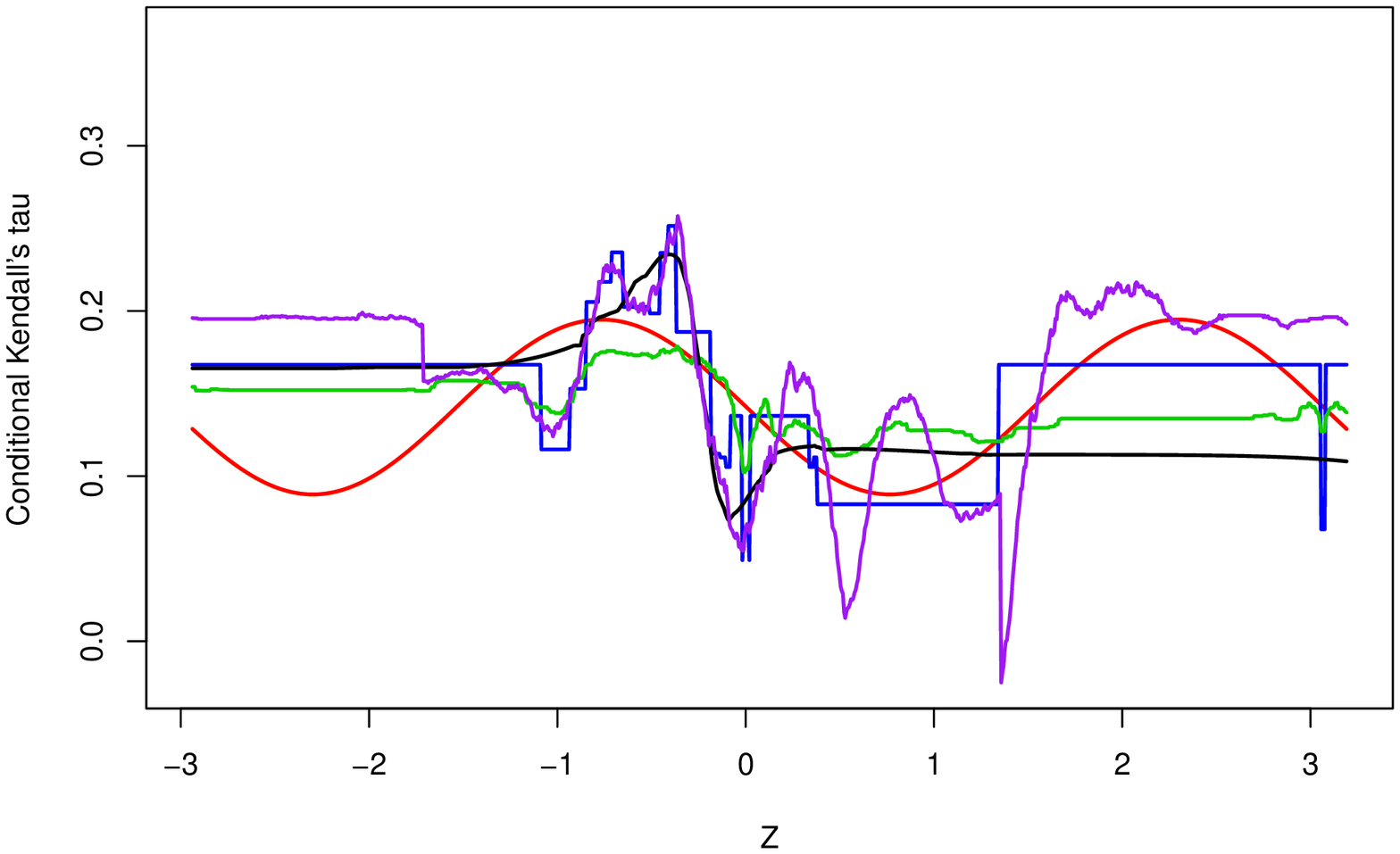}
        \caption{Conditional Kendall's tau between $(X_1, X_4)$ given $\Delta\sigma^I$ during the After-crisis period}
        \label{fig:CKT_X1_X4_Z2_P4}
    \end{minipage}
\end{figure} 

\subsection{Conditional dependence with respect to the variations $\Delta\sigma^I$ of the Eurostoxx's implied volatility index}

The implied volatility is computed using option prices. In this sense, this financial quantity reflects investors' anticipation of future uncertainty. When important events happen, investors most often update their anticipations, which results in a change of implied volatilities.
This change, denoted by $\Delta\sigma^I$ may be linked to variations of the conditional dependence between stock returns of different countries. Figures \ref{fig:CKT_X1_X2_Z2_P3} to~\ref{fig:CKT_X1_X4_Z2_P4} illustrate the variations of the conditional Kendall's tau between couples of stock returns with respect to the conditioning variable $\Delta\sigma^I$ during the two periods we study.

\mds

For each couple, the levels of the conditional Kendall's tau are higher during the European debt crisis than during the after-crisis period. This is coherent with our conclusions in the previous subsection. But here, conditional Kendall's taus look like concave functions of $\Delta\sigma^I$ during the crisis, while they exhibit ``double bumps'' features after the crisis.
During the crisis, when $\Delta\sigma^I$ is small in absolute value, implied volatilities do not change much and the dependence is in general higher than during big changes of the market implied volatility, i.e. when $|\Delta\sigma^I|$ is high (see Figures \ref{fig:CKT_X1_X3_Z2_P3} and \ref{fig:CKT_X1_X4_Z2_P3}).

\mds

One exception is the couple (France, Germany), for which the conditional Kendall's tau is roughly a decreasing function of $\Delta\sigma^I$ during the crisis. France and Germany are close countries and have strong economic relationships, but Germany is seen as a country in the ``center of Europe'' while France share a lot of similarities with countries of the periphery (in the South of Europe). Indeed, during the crisis, when implied volatility decreases (corresponding to a negative value of $\Delta\sigma^I$), the dependence is higher, which can be interpreted as investors seeing the two countries as close. On the contrary, when the market implied volatility increases, there are concerns in the market about the robustness of Eurozone and investors raise doubts about southern European countries - including France - which tends to decrease the conditional Kendall's tau between French and German returns.

\mds

After the crisis, the couples (Germany, France), and (Germany, Denmark) revert to a more usual shape of conditional dependence: when volatility does not change much, conditional Kendall's tau is low ; when volatility changes much, conditional Kendall's tau is higher, reflecting more stressed situations.
In this period, an exception is the couple (Germany, Greece), whose conditional Kendall's tau has a particular shape, that looks like the one of the couple (Germany, France) during the crisis. This is coherent with the fact that, in stressed situations, when volatility increases, investors sometimes remembers that Greece still has a fragile economy, which results in a lower conditional Kendall's tau. But three estimators suggest that, when volatility increases very much, conditional Kendall's tau between Germany and Greece increases again, following the classical tendencies that we had already observed.

\section{Conclusion}

In a parametric setting, we have proposed a localized log-likelihood method to estimate conditional Kendall's tau.  When the link function is analytically tractable and explicit, it is then possible to code and optimize the full penalized criterion. The consistency and the asymptotic normality of such estimators have been stated. In particular, this is the case for Logit or Probit-type link functions. We noticed that evaluating a Kendall's tau is equivalent to evaluating a probability of being classified as a concordant pair. Therefore, most classification procedures can be adapted to directly estimate conditional Kendall's tau. Classification trees, random forests, nearest neighbors and neural networks have been discussed. They generally provide more flexible parametric models than previously.

\mds

\begin{table}[htb]
    \centering
    \renewcommand{\arraystretch}{1.5}
    \resizebox{\textwidth}{!}{%
    \begin{tabular}{c|ccccc}
        \multirow{2}{*}{Method}
        & Performance
        & Computation
        & Interpretation 
        & \multicolumn{2}{c}{Tuning parameters} \\
        & in the sense of (\ref{eq:def:criteria_l2}) & time &
        & Number & Difficulty of choice \\
        \hline
        Logit / Probit (well-specified) & Best & Very Slow & Yes & 1 & Easy \\
        Logit / Probit (mis-specified) & Low & Very Slow & Possible & 1 & Easy \\
        Tree & Average & Very Fast & Possible & 3 (see \cite{ripley2018tree}) & Average \\
        Random forests & Good & Average & No & at least 4 & Average
        \\
        Nearest neighbors & Very Good & Fast & No & at least 5 & Complicated\\
        Neural network & Excellent & Slow & No & at least 2 & Complicated \\
        \hline
    \end{tabular}
    }
    \caption{Strengths and weaknesses of the proposed estimation procedures}
    \label{tab:strengths_weakneses_estimators}
\end{table}

We note that multiple trade-offs arise when choosing one of these methods, as displayed in Table~\ref{tab:strengths_weakneses_estimators}. Depending on the necessities of the situation, statisticians can choose algorithms that best match their needs.
To summarize, trees and random forests methods are the fastest ones, but exhibit the lowest performances. 
Parametric methods such as the logit and probit may be very performing under some ``simple'' functional forms of $g$ and $\psibm$, but they deteriorate quickly when the true underlying model departs from their parametric specification.
Note that they also show the longest computation time. Nonetheless, interpretability of the coefficient $\beta$ can be useful in applications. Even if the model is misspecified, it can still be seen as an estimation of the best approximation of $\z \mapsto \tau_{1,2|\Z=\z}$ on the function space generated by $\psibm$. Nearest neighbors methods are average in terms of computation time as well as performance. 
Finally, neural networks are the slowest of all our nonparametric methods, but they behave nearly uniformly the best ones in term of prediction.
Finally, we have evaluated these different methods on several empirical illustrations. 

\mds

\appendix

\section{Some basic definitions about copulas}
\label{reminder_copulas}

Here, we recall the main concepts around copulas and conditional copulas.
First, a $d$-dimensional copula is a cdf on $[0,1]^d$ whose margins are uniform distributions. 
Sklar's theorem states that, for any $d$-dimensional distributions $H$, whose marginal cdfs' are denoted as $F_1,\ldots,F_d$, there exists a copula $C$ s.t.
\begin{equation}
H(x_1,\ldots,x_d)=C\big( F_1(x_1),\ldots,F_d(x_d)  \big), 
\label{Sklar}
\end{equation}
for every $(x_1,\ldots,x_d)\in\Rb^d$. If the law of $H$ is continuous, the latter $C$ is unique, and it is called {\em the copula} associated to $H$.
Inversely, for a given copula and some univariate cdfs' $F_k$, $k=1,\ldots,d$, Equation~(\ref{Sklar}) defines a $d$-dimensional cdf $H$.

\mds

The latter concept of copula is similarly related to any random vector $\X$ whose cdf is $H$, and there is no ambiguity by using the same term. Copulas are invariant w.r.t. strictly increasing transforms of the margins $X_k$, $k=1,\ldots,d$. They provide very practical tools for modeling complex and/or highly dimensional distributions in a flexible way, by splitting the task into two parts: the specification of the marginal distributions on one side, and the specification of the copula on the other side. Therefore, a copula can be seen as a function that describes the dependence between the components of $\X$, 
independently of the marginal distributions. Several popular dependence measures are functionals of the underlying copula only: Kendall's tau, Spearman's rho, Blomqvist coefficient, etc. The classical textbooks by Joe~\cite{joeBook2015} or Nelsen~\cite{nelsen2007introduction} provide numerous and detailed results.

\mds

Numerous parametric families of copulas have been proposed in the literature: Gaussian, Student, Archimedean, Marshall-Olkin, extreme-value, etc.  
Several inference methods have been adapted to evaluate an underlying copula, possible without estimating the marginal cdfs' (Canonical Maximum Likelihood). 
See Cherubini and Luciano~\cite{cherubini} for details. Nonparametric methods have been developed too, since the seminal papers of Deheuvels~\cite{deheuvels1979,deheuvels1981} about empirical copula processes. 

\mds

Second, conditional copulas have been formally introduced by Patton~\cite{patton2006a,patton2006b}. They are rather straightforward extensions of the latter concepts, when dealing with conditional distributions. Formally, for a given sigma-algebra $\Fc$, let $H(\cdot \vert \Fc)$ (resp. $F_k(\cdot \vert \Fc)$) be the conditional distribution of $\X$ (resp. $X_k$, $k=1,\ldots,d$) given $\Fc$. 
The ``conditional version of'' Sklar's theorem now states that there exists a random copula $C(\cdot \vert \Fc)$ s.t.
\begin{equation}
H(x_1,\ldots,x_d \vert \Fc )=C\big( F_1(x_1\vert \Fc ),\ldots,F_d(x_d\vert \Fc ) \vert \Fc   \big), \;\; \text{a.e.} 
\label{Sklar_cond}
\end{equation}
for every $(x_1,\ldots,x_d)\in\Rb^d$. If the law of $H(\cdot \vert \Fc)$ is continuous, the latter $C(\cdot \vert \Fc )$ is unique, and it is called {\em the conditional copula} associated to $H(\cdot \vert \Fc )$, given $\Fc$.
Inversely, given $\Fc$, a conditional copula $C(\cdot \vert \Fc )$ and some univariate cdfs' $F_k(\cdot \vert \Fc )$, $k=1,\ldots,d$, Equation~(\ref{Sklar_cond}) defines a $d$-dimensional conditional cdf $H(\cdot \vert \Fc )$.
See Fermanian and Wegkamp~\cite{FermanianWegkamp} for extensions of the latter concepts.

\section{Proof of Theorem~\ref{consistency_beta}}
\label{proof_consistency}
Simple calculations provide: if $i\neq j$ and under~(\ref{model:cond_tau_Z}),
\begin{eqnarray*}
\lefteqn{ 
\EE[ L_n(\beta)]= \EE\left[ K_h(\Z_i - \Z_j) \ell_\beta (W_{(i,j)},\Z_i)  \right]  = \EE\left[  K_h(\Z_i - \Z_j) \EE[ \ell_\beta (W_{(i,j)},\Z_i) |\Z_i,\Z_j] \right]  }\\
&=&
\EE\left[ K_h(\Z_i - \Z_j) \bigg(
 p(\Z_i,\Z_j)  \log \left( \frac{1}{2} +
\frac{1}{2} g ( \psibm( \Z_{i})^T \beta) \right)
    + (1-p (\Z_i,\Z_j)) \log
    \left( \frac{1}{2} - \frac{1}{2} g ( \psibm( \Z_i)^T \beta) \right) \bigg) \right] \\
&=& \EE\left[ K_h(\Z_i - \Z_j) \phi(\Z_i,\Z_j,\beta) \right] \\
&=& \EE\left[ \int K(\t) \phi(\Z_i,\Z_i-h\t,\beta) f_{\Z}(\Z_i-h\t) \, d\t \right],
\end{eqnarray*}
that tends to $\EE\left[  \phi(\Z_i,\Z_i,\beta) f_{\Z}(\Z_i)\right]=L_\infty (\beta)$ when $n\rightarrow \infty$, if
$ \int \big( \phi(\z,\cdot,\beta) f_\Z(\cdot) \big)_\eps (\z) f_{\z}(\z)  \, d\z  <\infty,$
for some $\eps>0$ (invoke the dominated convergence Theorem and the compact support of $K$).

\mds

Now, let us prove that, for any $\beta$, $L_n(\beta)$ tends towards $L_{\infty}(\beta)$ in probability, when $n\rightarrow \infty$.
It is sufficient to prove that the variance of $L_n(\beta)$ tends to zero.
\begin{eqnarray*}
\lefteqn{ \EE\Big[ \bigg(  L_n(\beta) - \EE[L_n(\beta)] \bigg)^2 \Big] =
\frac{1}{n^2(n-1)^2}\sum_{i_1,j_1;i_1\neq j_1} \sum_{i_2,j_2;i_2\neq j_2}   }\\
& & \Big( \EE\left[ K_h(\Z_{i_1} - \Z_{j_1}) K_h(\Z_{i_2} - \Z_{j_2}) \ell_\beta (W_{(i_1,j_1)},\Z_{i_1})  \ell_\beta (W_{(i_2,j_2)},\Z_{i_2})    \right]
- \EE[ L_n(\beta)  ]^2   \big) \\
&=:&
\frac{1}{n^2(n-1)^2}\sum_{i_1,j_1;i_1\neq j_1} \sum_{i_2,j_2;i_2\neq j_2} v_{i_1,j_1,i_2,j_2},\hspace{5cm}
\end{eqnarray*}
with obvious notations. Obviously, $v_{i_1,j_1,i_2,j_2}$ is zero when $i_1$ and $j_1$ are not equal to $i_2$ nor $j_2$.
At the opposite, in the case of equalities between some of these four indices, we get non-zero terms.

\mds

To be specific, when $i_1=i_2=i$, and $j_1\neq j_2$, we have
\begin{eqnarray*}
\lefteqn{v_{i,j_1,i,j_2} = \EE\left[ K_h(\Z_{i} - \Z_{j_1}) K_h(\Z_{i} - \Z_{j_2}) \ell_\beta (W_{(i,j_1)},\Z_{i})  \ell_\beta (W_{(i,j_2)},\Z_{i})\right]- \EE[ L_n(\beta)  ]^2    }\\
&=& \EE\left[
\int  K(\x) K(\y) A(\Z_i, \Z_i-h\x,\Z_i-h\y)  f_\Z (\Z_i-h\x) f_\Z (\Z_i-h\y)\,  d\x \, d\y \right]- \EE[ L_n(\beta)  ]^2,
\end{eqnarray*}
by setting
\begin{eqnarray*}
\lefteqn{A(\x,\y,\z) :=\EE\left[ \ell_\beta (W_{(i,j_1)},\Z_{i})  \ell_\beta (W_{(i,j_2)},\Z_{i}) | \Z_i=\x,\Z_{j_1}=\y,\Z_{j_2}=\z \right]       }\\
&= & p(\x,\y,\z) \log^2 q(\x,\beta) + (p(\x,\y) - p(\x,\y,\z)) \log q(\x,\beta) \log (1- q(\x,\beta))  \\
&+& (p(\x,\z)- p(\x,\y,\z)) \log q(\x,\beta) \log (1- q(\x,\beta)) + (1- p(\x,\y) -p(\x,\z)+p(\x,\y,\z))  \log^2 (1-q(\x,\beta)).
\end{eqnarray*}
If  $ \int A(\z, \cdot, \cdot)_\eps(\z,\z)  f^2_{\Z,\eps}(\z) f_\Z(\z)\, d\z <\infty  $ for some $\eps>0$, then $v_{i,j_1,i,j_2}$ tends to a constant when $n\rightarrow \infty$ (independently of the choice of such indices).
\mds

A similar analysis can be led for the other terms.
When $i_1=j_2$ and $j_1\neq i_2$, we get
\begin{eqnarray*}
\lefteqn{v_{i_1,j_1,i_2,i_1} = \EE\left[ K_h(\Z_{i_1} - \Z_{j_1}) K_h(\Z_{i_2} - \Z_{i_1}) \ell_\beta (W_{(i_1,j_1)},\Z_{i_1})
\ell_\beta (W_{(i_2,i_1)},\Z_{i_2})\right]- \EE[ L_n(\beta)  ]^2    }\\
&=& \EE\left[ 
\int  K(\x) K(\y) B(\Z_{i_1},\Z_{i_1}-h\x,\Z_{i_1}+h\y) f_\Z (\Z_{i_1}-h\x) f_\Z (\Z_{i_1}+h\y)\,  d\x \, d\y \right]- \EE[ L_n(\beta)  ]^2,
\end{eqnarray*}
by setting
\begin{eqnarray*}
\lefteqn{B(\x,\y,\z) :=\EE\left[ \ell_\beta (W_{(i_1,j_1)},\Z_{i_1})  \ell_\beta (W_{(i_2,i_1)},\Z_{i_2}) | \Z_{i_1}=\x,\Z_{j_1}=\y,\Z_{i_2}=\z \right]       }\\
&= & p(\x,\y,\z) \log q(\x,\beta)\log q(\z,\beta) + (p(\x,\y) - p(\x,\y,\z)) \log q(\x,\beta) \log (1- q(\z,\beta))  \\
&+& (p(\x,\z)- p(\x,\y,\z)) \log q(\z,\beta) \log (1- q(\x,\beta)) \\
&+& (1- p(\x,\y) -p(\x,\z)+p(\x,\y,\z))  \log (1-q(\x,\beta))\log (1-q(\z,\beta)).
\end{eqnarray*}
If  $ \int B(\z, \cdot, \cdot)_\eps(\z,\z) f^2_{\Z,\eps}(\z)  f_{\Z}(\z)\, d\z <\infty  $, then $v_{i_1,j_1,i,i_1}$ tends to a constant when $n\rightarrow \infty$.

\mds

When $j_1=j_2=j$ and $i_1\neq i_2$, we obtain
\begin{eqnarray*}
\lefteqn{v_{i_1,j_1,i_2,j_2} = \EE\left[ K_h(\Z_{i_1} - \Z_{j}) K_h(\Z_{i_2} - \Z_{j}) \ell_\beta (W_{(i_1,j)},\Z_{i_1})
\ell_\beta (W_{(i_2,j)},\Z_{i_2})\right]- \EE[ L_n(\beta)  ]^2    }\\
&=& \EE\left[ 
\int K(\x) K(\y) C(\Z_j+h\x,\Z_{j},\Z_{j}+h\y) f_\Z (\Z_j+h\x) f_\Z (\Z_j +h\y)\,  d\x \, d\y \right]- \EE[ L_n(\beta)  ]^2,
\end{eqnarray*}
by setting
\begin{eqnarray*}
\lefteqn{C(\x,\y,\z) :=\EE\left[ \ell_\beta (W_{(i_1,j)},\Z_{i_1})  \ell_\beta (W_{(i_2,j)},\Z_{i_2}) | \Z_{i_1}=\x,\Z_{j}=\y,\Z_{i_2}=\z \right]       }\\
&= & p(\x,\y,\z) \log q(\x,\beta)\log q(\z,\beta) + (p(\x,\y) - p(\x,\y,\z)) \log q(\x,\beta) \log (1- q(\z,\beta))  \\
&+& (p(\y,\z)- p(\x,\y,\z)) \log q(\z,\beta) \log (1- q(\x,\beta)) \\
&+& (1- p(\x,\y) -p(\y,\z)+p(\x,\y,\z))  \log (1-q(\x,\beta))\log (1-q(\z,\beta)) .
\end{eqnarray*}
If  $ \int C( \cdot,\z, \cdot)_\eps(\z,\z) f^2_{\Z,\eps}(\z)  f_{\Z}(\z)\, d\z <\infty  $, then $v_{i_1,j,i_2,j}$ tends to a constant when $n\rightarrow \infty$.

\mds

When $j_1=i_2$ and $i_1\neq j_2$:
\begin{eqnarray*}
\lefteqn{v_{i_1,j_1,j_1,j_2} = \EE\left[ K_h(\Z_{i_1} - \Z_{j_1}) K_h(\Z_{j_1} - \Z_{j_2}) \ell_\beta (W_{(i_1,j_1)},\Z_{i_1})
\ell_\beta (W_{(j_1,j_2)},\Z_{j_1})\right]- \EE[ L_n(\beta)  ]^2    }\\
&=& \EE\left[
\int  K(\x) K(\y) D(\Z_{j_1}+h\x,\Z_{j_1},\Z_{j_1}-h\y) f_\Z (\Z_{j_1}+h\x) f_\Z (\Z_{j_1}-h\y)\,  d\x \, d\y \right]- \EE[ L_n(\beta)  ]^2,
\end{eqnarray*}
by setting
\begin{eqnarray*}
\lefteqn{D(\x,\y,\z) :=\EE\left[ \ell_\beta (W_{(i_1,j_1)},\Z_{i_1})  \ell_\beta (W_{(j_1,j_2)},\Z_{j_1}) | \Z_{i_1}=\x,\Z_{j_1}=\y,\Z_{j_2}=\z \right]       }\\
&= & p(\x,\y,\z) \log q(\x,\beta)\log q(\y,\beta) + (p(\x,\y) - p(\x,\y,\z)) \log q(\x,\beta) \log (1- q(\y,\beta))  \\
&+& (p(\y,\z)- p(\x,\y,\z)) \log q(\y,\beta) \log (1- q(\x,\beta)) \\
&+& (1- p(\x,\y) -p(\y,\z)+p(\x,\y,\z))  \log (1-q(\x,\beta))\log (1-q(\y,\beta)) .
\end{eqnarray*}
If  $ \int D( \cdot,\z, \cdot)_\eps(\z,\z) f^2_{\Z,\eps}(\z) f_{\Z}(\z)\, d\z <\infty  $, then $v_{i_1,j_1,j_1,j_2}$ tends to a constant when $n\rightarrow \infty$.

\mds

There are two cases of two equalities.
If $i_1=i_2=i$ and $j_1=j_2=j$, this yields
\begin{eqnarray*}
\lefteqn{v_{i,j,i,j} = \EE\left[ K_h(\Z_{i} - \Z_{j})^2 \ell^2_\beta (W_{(i,j)},\Z_{i})\right]- \EE[ L_n(\beta)  ]^2    }\\
&=& h^{-p} \EE\left[
\int  K(\x)^2 E(\Z_i,\Z_{i}-h\x) f_\Z (\Z_i-h\x) \,  d\x \right] - \EE[ L_n(\beta)  ]^2,
\end{eqnarray*}
by setting
\begin{eqnarray*}
\lefteqn{E(\x,\y) :=\EE\left[ \ell^2_\beta (W_{(i,j)},\Z_{i}) | \Z_{i}=\x,\Z_{j}=\y \right]       }\\
&= & p(\x,\y) \log^2 q(\x,\beta) + (1- p(\x,\y))  \log^2 (1-q(\x,\beta)) .
\end{eqnarray*}
If  $ \int E( \z,\cdot)_\eps(\z)  f_{\Z,\eps}(\z)  f_{\Z}(\z)\, d\z <\infty  $, then $h^p v_{i,j,i,j}$ tends to a constant when $n\rightarrow \infty$.

\mds

Finally, if $i_1=j_2$ and $j_1=i_2$, we get
\begin{eqnarray*}
\lefteqn{v_{i_1,j_1,i_1,j_1} = \EE\left[ K_h(\Z_{i_1} - \Z_{j_1})^2 \ell_\beta (W_{(i_1,j_1)},\Z_{i_1}) \ell_\beta (W_{(j_1,i_1)},\Z_{j_1})\right]- \EE[ L_n(\beta)  ]^2    }\\
&=& h^{-p} \EE\left[ 
\int  K(\x)^2 F(\Z_i,\Z_{i}-h\x) f_\Z (\Z_i-h\x) \,  d\x \right]- \EE[ L_n(\beta)  ]^2,
\end{eqnarray*}
by setting
\begin{eqnarray*}
\lefteqn{F(\x,\y) :=\EE\left[ \ell_\beta (W_{(i_1,j_1)},\Z_{i_1}) \ell_\beta (W_{(j_1,i_1)},\Z_{j_1})  | \Z_{i_1}=\x,\Z_{j_1}=\y \right]       }\\
&= & p(\x,\y) \log q(\x,\beta)\log q(\y,\beta) + (1- p(\x,\y))  \log (1-q(\x,\beta)) \log (1-q(\y,\beta)).
\end{eqnarray*}
If  $ \int F( \z,\cdot)_\eps(\z)  f_{\Z,\eps}(\z) f_{\Z}(\z)\, d\z <\infty  $, then $h^p v_{i_1,j_1,j_1,i_1}$ tends to a constant when $n\rightarrow \infty$.

\mds

Summarizing the previous terms, we have obtained that $Var(L_n(\beta))= O(n^{-1}+ n^{-2}h^{-p})$, that tends to zero pointwise, when $n^2h^p \rightarrow \infty$.
We deduce $L_n(\beta)-L_{\infty}(\beta)= L_n(\beta) - \EE[L_n(\beta)] + \EE[L_n(\beta)]-L_{\infty}(\beta) = o_P(1)$.
Since $L_n(\cdot)$ and $L_{\infty}(\cdot)$ are concave, invoking the convexity lemma of Geyer \cite{geyer1996} (see Knight and Fu \cite{knight2000asymptotics}, alternatively),
the maximizer $\hat\beta$ of $L_n$ tends in probability towards the maximizer of $L_\infty $. $\Box$

\mds 

We summarize the latter technical assumptions that are sufficient to obtain the consistency of $\hat\beta$: for some $\eps>0$,
\begin{equation}
 \int \big( \phi(\z,\cdot,\beta) f_\Z(\cdot) \big)_\eps (\z) f_{\z}(\z)  \, d\z  <\infty,
\label{cond_reg_consistency}
\end{equation}

\begin{equation}
 \int \Big( A(\z, \cdot, \cdot)_\eps(\z,\z) + B(\z, \cdot, \cdot)_\eps(\z,\z) +C( \cdot,\z, \cdot)_\eps(\z,\z)  + D( \cdot,\z, \cdot)_\eps(\z,\z)   \Big)
 f^2_{\Z,\eps}(\z) f_\Z(\z)\, d\z <\infty  ,
\label{cond_consis_1}
\end{equation} 
\begin{equation}
\int \Big( \phi(\z,\cdot,\beta)_\eps (\z) + E( \z,\cdot)_\eps(\z) + F( \z,\cdot)_\eps(\z) \Big) f_{\Z,\eps}(\z)  f_{\Z}(\z)\, d\z <\infty  .
 \label{cond_consis_2}
\end{equation}

\section{Proof of Theorem~\ref{AN_beta}}
\label{proof_AN}

Set $\u:=\sqrt{n}(\beta- \beta^*)$ and $\hat\u:=\sqrt{n}(\hat\beta- \beta^*)$. Obviously,
\begin{eqnarray*}
\lefteqn{ \hat \u = \arg \max_{\u \in \Rb^{p'}} L_n(\beta^* + n^{-1/2}\u) - \lambda_n |\beta^* + n^{-1/2}\u|_1,
  }\\
&= &  \arg \max_{\u \in \Rb^{p'}} nL_n(\beta^* + n^{-1/2}\u) - n L_n(\beta^*) -n \lambda_n \big\{ |\beta^* + n^{-1/2}\u|_1-|\beta^*|_1     \big\}.
\end{eqnarray*}
Note that
\begin{eqnarray*}
\lefteqn{  n\lambda_n |\beta^* + n^{-1/2}\u|_1-n|\beta^*|_1 = n^{1/2}\lambda_n \sum_{k;\beta_k^*=0}  |u_k| +
 n^{1/2}\lambda_n \sum_{k;\beta_k^* \neq 0}  sign(\beta_k^*) u_k   }\\
&\longrightarrow & \mu \sum_{k;\beta_k^*=0} |u_k| + \mu\sum_{k;\beta_k^* \neq 0} sign(\beta_k^*) u_k , \hspace{4cm}
\end{eqnarray*}
when $n\rightarrow \infty$. 
Moreover, 
$$ nL_n(\beta^* + n^{-1/2}\u) - n L_n(\beta^*)= n^{1/2} \dot L_n(\beta^*).\u + \frac{1}{2}\u^T \ddot{L}_n(\bar\beta) \u+ \frac{1}{6\sqrt{n}}\dddot{L}_n(\bar\beta).\u^{(3)},$$
for some (random) $\bar\beta$ s.t. $|\beta^* - \bar\beta| < |\beta^* - \beta|$.
We will successively prove that
\begin{itemize}
\item[(i)] $n^{1/2}\dot L_n(\beta^*)$ weakly tends to a Gaussian random vector $ \Wb$, $\Wb\sim \Nc(0_p,\Sigma_{\beta^*})$;
\item[(ii)] $\ddot{L}_n(\beta^*)$ tends in probability towards a constant matrix $\Hb(\beta^*)$;
\item[(iii)] $\dddot{L}_n(\bar\beta)$ is $O_P(1)$. 
\end{itemize}
Then, $\hat\u =   \arg \max_{\u} \Lc_n(\u)$, where $\Lc_n(\u)$ weakly tends to  
$$ \Lc_\infty(\u):= \Wb .\u+ \frac{1}{2}\u^T \Hb(\beta^*) \u-
\mu \sum_{k;\beta_k^*=0} |u_k| - \mu\sum_{k;\beta_k^* \neq 0} sign(\beta_k^*) u_k ,$$
that is concave. 
The result will follow, applying Theorem 1 in Kato \cite{kato2009asymptotics}. 

\mds

First, let us prove (i), i.e. the asymptotic normality of $n^{1/2}\dot L_n(\beta)$ for any given parameter $\beta$.
Consider the centered criterion
$$ M_n(\beta) := \dot L_n(\beta) - \EE[\dot L_n(\beta)]= \frac{1}{n(n-1)} \sum_{i,j;i\neq j}  \ell_{ij}(\beta) ,$$
where $\ell_{ij} :=K_h(\Z_i - \Z_j) \partial_\beta \ell_\beta( W_{(i,j)}, \Z_i) - \EE[\dot L_n(\beta)]$. We symmetrize the localized likelihood:
$$ M_n(\beta)= \frac{1}{2n(n-1)} \sum_{i,j;i\neq j}  M_{ij}(\beta) ,$$
 where $M_{ij}(\beta)$ (or simply $M_{ij}$) is $\ell_{ij}(\beta)+ \ell_{ji}(\beta)$. Note that $M_{ij}=M_{ji}$ and that $\EE[M_{ij}]=0$. 

\mds

By the dominated convergence theorem and a change of variable, we easily check that $\EE[ \dot L_n(\beta)]=\partial_\beta 
L_\infty (\beta) + o(1)$ if, for some $\eps>0$, we have
 $\int \big( \partial_\beta\phi(\z,\cdot,\beta) f_\Z(\cdot) \big)_\eps (\z) f_{\z}(\z)  \, d\z  <\infty$.
Moreover, by simple calculations, we get, if $i\neq j$,
\begin{eqnarray*}
\lefteqn{ \EE [ M_{n} | \Z_i]= \frac{1}{2n}  \EE[M_{ij} + M_{ji} | \Z_i]=\frac{1}{n}  \EE[M_{ij} | \Z_i]     }\\
&=& \frac{1}{n} \int \{ K_h(\Z_i - \z)  \partial_\beta\phi(\Z_i,\z,\beta) + K_h( \z- \Z_i) \partial_\beta\phi(\z,\Z_i,\beta)  \} f_\Z(\z) \, d\z - \frac{2}{n}\EE[\dot L_n(\beta)] \\
&=& \frac{1}{n} \int  K(\t) \{  \partial_\beta\phi(\Z_i,\Z_i-h\t,\beta) f_\Z(\Z_i-h\t)+ \partial_\beta\phi(\Z_i+h\t,\Z_i,\beta) f_\Z(\Z_i+h\t) \}  \, d\t - \frac{2}{n}\EE[\dot L_n(\beta)]\\
&=& \frac{2}{n} \partial_\beta\phi(\Z_i,\Z_i,\beta) f_\Z(\Z_i)-\frac{2}{n}\dot L_\infty(\beta) + o(n^{-1}) +r_{n,i},
\end{eqnarray*}
where, by a $m$-order limited expansion, we obtain 
$$ \|r_{n,i}\| \leq \frac{Cst. h^m \int |K|}{n m!}  \|  (f_\Z(\cdot) \partial_\beta\phi(\Z_i, \cdot,\beta))^{(m)} 
+ (f_\Z(\cdot)\partial_\beta\phi( \cdot,\Z_i,\beta))^{(m)}\|_\eps(\Z_i),  $$
for any norm $\|\cdot \|$ on $\Rb^p$ and a positive constant $Cst$. 
We deduce that $n^{1/2}\sum_{i=1}^n \EE[ M_{n} | \Z_i]$ is asymptotically normal by invoking the usual CLT, under condition~(\ref{cond_reg_fZphi}) below. 
To be specific, the H\'ajek projection of $M_n$ is
\begin{equation*}
\frac{\sqrt{n}}{2}\sum_{i=1}^n \EE [ M_{n} | \Z_i] = 
\frac{1}{\sqrt{n}} \sum_{i=1}^n \{ \partial_\beta\phi(\Z_i,\Z_i,\beta) f_\Z(\Z_i)-\dot L_\infty(\beta) \} +o_P(1)\leadsto \Nc (0,\Sigma_\beta), \;\text{with}
\end{equation*}
$$ \Sigma_\beta := \int \partial_\beta\phi(\z,\z,\beta)  \partial_{\beta}\phi(\z,\z,\beta)^T f^3_\Z(\z) \, d\z -\dot L_\infty(\beta)\dot L_\infty(\beta)^T.$$
Note that that $\dot L_\infty(\beta^*)=0$.

\mds
  
Now consider the ``remainder term'' 
$\Delta_n:= M_n (\beta) - \sum_{i=1}^n \EE[M_n |\Z_i]/2$.
Since $\EE[M_n |\Z_i]= n^{-1}\EE[M_{ij} | \Z_i]$, we deduce
\begin{equation*} 
\Delta_n= M_n (\beta) - \frac{1}{2}\sum_{i=1}^n \EE[M_n |\Z_i]=  \frac{1}{2n(n-1)} \sum_{i,j;i \neq j}  \{ M_{ij} - \EE[M_{ij} |\Z_i]\}.
\end{equation*}
It is relatively easy to prove that $\Delta_n$ is negligible, i.e. $\Delta_n=o_P(n^{-1/2})$. 
Indeed, let us prove that the variance of $n^{1/2}\Delta_n$ tends to zero with $n$:
$$ Var(n^{1/2}\Delta_n)=n\EE[\Delta_n \Delta_n^T]= \frac{1}{4n(n-1)^2} \sum_{i_1,j_1;i_1 \neq j_1} \sum_{i_2,j_2;i_2 \neq j_2} \delta(i_1,i_2,j_1,j_2),$$
$$ \delta(i_1,i_2,j_1,j_2) = \EE\left[  \{ M_{i_1 j_1} - \EE[M_{i_1 j_1} |\Z_{i_1}]\} \times \{ M_{i_2 j_2} - \EE[M_{i_2 j_2} |\Z_{i_2}]\}^T \right]   .$$
If there is no identity among the indices $(i_1,j_1,i_2,j_2)$, with $i_1\neq j_1$ and $i_2\neq j_2$, then $\delta(i_1,i_2,j_1,j_2)$ is zero. 
Moreover, this is still the case when there is only a single identity. For instance, assume $i_1=i_2=i$ and $j_1\neq j_2$. Then, 
 \begin{eqnarray*}
\lefteqn{ \delta(i,i,j_1,j_2) = \EE\left[  \{ M_{i j_1} - \EE[M_{i j_1} |\Z_{i}]\} \times \{ M_{i j_2} - \EE[M_{i j_2} |\Z_{i}]\}^T \right]  }\\
&= & \EE\left[  \{ M_{i j_1} - \EE[M_{i j_1} |\Z_{i}]\} \times \EE[ \{ M_{i j_2} - \EE[M_{i j_2} |\Z_{i}]\}^T |\Z_i,\Z_{j_1}] \right]  \\
&= & \EE\left[  \{ M_{i j_1} - \EE[M_{i j_1} |\Z_{i}]\} \times 0 \right] =0.
\end{eqnarray*}
The other terms for which a single identity between the indices can be managed similarly. 

\mds

At the opposite, non-zero terms appear when $i_1=i_2=i$ and $j_1=j_2=j$. In this case, we obtain
$$ \delta(i,i,j,j) = \EE\left[  \{ M_{i j} - \EE[M_{i j} |\Z_{i}]\} \{ M_{i j} - \EE[M_{i j} |\Z_{i}]\}^T \right] =\EE[  M_{i j} M_{i j}^T ] 
- \EE[ \EE[M_{i j} |\Z_{i}] \EE[M_{i j} |\Z_{i}]^T ] .$$
By a usual change of variable and by symmetry, we get
\begin{eqnarray*}
\lefteqn{ \EE[  M_{i j} M_{i j}^T ] = 2\EE\left[  \{ \ell_{ij}(\beta)\ell_{ij}(\beta)^T + \ell_{ij}(\beta) \ell_{ji}(\beta)^T \}       \right]     }\\
&=& 2 \EE\left[ K_h^2(\Z_i-\Z_j)  \partial_\beta\ell_\beta(W_{(i,j)},\Z_i) \partial_\beta \ell_\beta(W_{(i,j)},\Z_i)^T \right. \\
&  + & \left. K_h(\Z_i-\Z_j)K_h(\Z_j-\Z_i) \partial_\beta\ell_\beta(W_{(i,j)},\Z_i) \partial_\beta\ell_\beta(W_{(j,i)},\Z_j)^T \right] + O(1) \\
&=& 2h^{-p} \EE\left[ \int \big( K^2(\x)    H_1(\Z_i,\Z_i-h\x)+ K(\x)K(-\x) H_2(\Z_i,\Z_i - h\x) \big) f_\Z (\Z_i-h\x)\, d\x \right] + O(1)\\
&=& 2h^{-p} \int \big( K^2(\x)    H_1(\z,\z-h\x)+ K(\x)K(-\x) H_2(\z,\z - h\x) \big) f_\Z (\z-h\x) f_\Z (\z)\, d\x\, d\z + O(1),
\end{eqnarray*}
by setting
\begin{eqnarray*}
\lefteqn{ H_1(\x,\y)  = \EE\left[  \partial_\beta\ell_\beta(W_{(i,j)},\Z_i) \partial_\beta \ell_\beta(W_{(i,j)},\Z_i)^T |\Z_i=\x,\Z_j=\y \right]   }\\
&=& p(\x,\y) \frac{\psibm(\x)\psibm(\x)^T g'(\psibm(\x)^T\beta)^2 }{(1+ g(\psibm(\x)^T\beta))^2} 
+ (1-p(\x,\y)) \frac{\psibm(\x)\psibm(\x)^T g'(\psibm(\x)^T\beta)^2 }{(1- g(\psibm(\x)^T\beta))^2},\;\text{and}
\end{eqnarray*}
\begin{eqnarray*}
\lefteqn{ H_2(\x,\y)  = \EE\left[  \partial_\beta\ell_\beta(W_{(i,j)},\Z_i) \partial_\beta \ell_\beta(W_{(j,i)},\Z_j)^T |\Z_i=\x,\Z_j=\y \right]   }\\
&=& p(\x,\y) \frac{\psibm(\x)\psibm(\y)^T g'(\psibm(\x)^T\beta) g'(\psibm(\y)^T\beta) }{(1+ g(\psibm(\x)^T\beta))(1+ g(\psibm(\y)^T\beta))}
+ (1-p(\x,\y)) \frac{\psibm(\x)\psibm(\y)^T g'(\psibm(\x)^T\beta)g'(\psibm(\y)^T\beta) }{(1- g(\psibm(\x)^T\beta))(1+ g(\psibm(\y)^T\beta))}\cdot
\end{eqnarray*}

Therefore, $\EE[  M_{i j}M_{i j}^T ]$ is $O(h^{-p})$, if $\int (\|H_1\| + \| H_2\|)_\eps(\z,\cdot) f_\Z(\cdot)_\eps(\z)  f_\Z (\z)\, d\z <\infty$.

\mds

The last possible case providing non-zero $\delta(i_1,i_2,j_1,j_2)$ is $i_1=j_2=i$ and $j_1=i_2=j$. Then, we obtain
$$ \delta(i,j,j,i) = \EE\left[  \{ M_{i j} - \EE[M_{i j} |\Z_{i}]\}  \times \{ M_{ j i} - \EE[M_{ j i} |\Z_{j}]\}^T \right] 
=\delta(i,i,j,j) ,$$
due to the symmetry of $M_{ij}$. 
Thus, we have proved that $Var(n^{1/2}\Delta_n)=O(n^{-1}h^{-p}))=o(1)$, which implies 
$$  n^{1/2}M_n (\beta) = \frac{n^{1/2}}{2}\sum_{i=1}^n \EE[M_n |\Z_i] + o_P(1).$$
We deduce that $n^{1/2}M_n (\beta)$ weakly tends towards the Gaussian random vector $\Nc(0_p,\Sigma_\beta)$ for any $\beta$. When $\beta=\beta^*$, 
$\partial_\beta L_\infty (\beta^*)=0$, and this yields (i).

\mds
  


Second, let us deal with (ii) above. It is easy to prove that $\ddot L_n(\beta^*)$ tends to $\Hb(\beta^*)$ in probability, when $n$ tends to the infinity.
Indeed, the arguments are exactly the same as in~\ref{proof_consistency}, where we have proved that $L_n(\beta^*)$ is convergent in probability. 
We only have to replace $\ell_\beta(\cdot,\cdot)$ by its second derivatives w.r.t. $\beta$. To save space, the specific derivations of such conditions of regularity are left to the reader: simply replace the functions $A$, $B$,...,$F$ of~\ref{proof_consistency} by their second derivatives w.r.t. $\beta$, taken at $\beta=\beta^*$, and rewrite~(\ref{cond_consis_1}) and~(\ref{cond_consis_2}).

\mds

Third, to prove (iii), it is sufficient to state that $\EE[\| \dddot L_n(\beta) \|]$ is bounded from above, uniformly w.r.t. $\beta$ in a small neighborhood of $\beta^*$.
By derivation, we get, for every indices $a,b,c$ in $\{1,\ldots,p'\}$,
\begin{eqnarray*}
    \lefteqn{ \EE[ | \frac{\partial^3}{\partial \beta_a\partial \beta_b \partial \beta_c  } L_n(\beta) | ] }\\
    &\leq & Cst\times 
    \EE\bigg[ |K|_h(\Z_i - \Z_j) \big\{
    p(\Z_i,\Z_j)  H(1,\Z_i,\beta,a,b,c,)
    + (1-p (\Z_i,\Z_j)) H(-1,\Z_i,\beta,a,b,c)  \big\} \bigg] ,
\end{eqnarray*}
where, for every $\delta\in \{1,-1\}$, 
\begin{eqnarray*}
     H(\delta,\Z_i,\beta,a,b,c) = \bigg(
    \frac{|g'( \psibm( \Z_i)^T \beta)|^3}{|1+\delta g ( \psibm( \Z_i)^T \beta)|^3}
    +
    \frac{|g'g''( \psibm( \Z_i)^T \beta)|}{|1+\delta g ( \psibm( \Z_i)^T \beta)|^2}+
    \frac{|g'''( \psibm( \Z_i)^T \beta)|}{|1+\delta  g ( \psibm( \Z_i)^T \beta)|} \bigg)
    | \psibm( \Z_i)_a   \psibm( \Z_i)_b\psibm( \Z_i)_c |.
\end{eqnarray*}
and $Cst$ denotes a real constant that depend on $g$ and $(a,b,c)$ only. 
Therefore, it is sufficient to assume that
\begin{equation*}
    \int  |K|(\t) \bigg( p(\,\z-h\t)  H(1,\z,\beta,a,b,c)
    + (1-p (\z,\z-h\t)) H(-1,\z,\beta,a,b,c)  \bigg) 
    f_\Z(\z)f_\Z(\z-h\t) \, d\t\, d\z  <\infty.
\end{equation*}
This is guaranteed by Assumption~(\ref{cond_reg_abc}) below.
Then, under the latter assumption, (iii) is stated and this finishes the proof. $\Box$

\mds

For convenience, let us gather the main technical assumptions that have been requested to prove Theorem~\ref{AN_beta}: for some $\eps>0$,
\begin{equation}
 \int \big( \partial_\beta\phi(\z,\cdot,\beta) f_\Z(\cdot) \big)_\eps (\z) f_{\z}(\z)  \, d\z  <\infty.
\label{cond_reg_AN_phi}
\end{equation}
\begin{equation}
 \EE \left[ \|  (f_\Z(\cdot) \partial_\beta\phi(\Z_i, \cdot,\beta))^{(m)} + (f_\Z(\cdot)\partial_\beta\phi( \cdot,\Z_i,\beta))^{(m)}\|_\eps(\Z_i)  \right]<\infty.
\label{cond_reg_fZphi}
\end{equation}
\begin{equation}
\int (\|H_1\| + \| H_2\|)_\eps(\z,\cdot) f_\Z(\cdot)_\eps(\z)  f_\Z (\z)\, d\z <\infty.
\label{cond_reg_H1H2}
\end{equation} 
For every indices $(a,b,c)\in \{1,\ldots,p'\}$ and for $\Vc(\beta^*)$, some (small) neighborhood around $\beta^*$,
\begin{equation}
\sup_{\beta \in \Vc(\beta^*)} \int  \bigg( (p(\z,\cdot)f_\Z(\cdot))_\eps (\z) H(1,\z,\beta,a,b,c)
    + ((1-p )(\z,\cdot)f_\Z(\cdot))_\eps (\z) H(-1,\z,\beta,a,b,c,)  \bigg) f_\Z(\z)\, d\z  < \infty.
\label{cond_reg_abc}
\end{equation}

\begin{rem}

Note that $||p(\cdot, \cdot)||_\infty \leq 1$. If $g$ and its derivatives are bounded,  Condition~(\ref{cond_reg_abc}) is satisfied if
\begin{equation*}
    \sup_{\beta \in \Vc(\beta^*)}   \sup_{\delta \in \{-1,1\}} \int \|\psibm(\z)\|^3 |1+\delta g ( \psibm(\z)^T \beta)|^{-3} f_{\Z,\eps}(\z) f_\Z(\z)\, d\z  < \infty.
\end{equation*}
\end{rem}

\section*{Acknowledgments}

This work is supported by the Labex Ecodec under the grant ANR-11-LABEX-0047 from the French Agence Nationale de la Recherche. The first author thanks Solt Kovács for inspiring ideas and a discussion which lead to this article.

\section*{References}


\bibliographystyle{elsarticle-num}

\bibliography{biblio}

\begin{thebibliography}{10}
\expandafter\ifx\csname url\endcsname\relax
  \def\url#1{\texttt{#1}}\fi
\expandafter\ifx\csname urlprefix\endcsname\relax\def\urlprefix{URL }\fi
\expandafter\ifx\csname href\endcsname\relax
  \def\href#1#2{#2} \def\path#1{#1}\fi

\bibitem{joeBook2015}
H.~Joe, Dependence Modeling with copulas, Chapman \& Hall, 2015.

\bibitem{nelsen2007introduction}
R.~B. Nelsen, An introduction to copulas, Springer Science \& Business Media,
  2007.

\bibitem{gijbels2011conditional}
I.~Gijbels, N.~Veraverbeke, M.~Omelka, Conditional copulas, association
  measures and their applications, Comput. Statist. Data Anal. 55~(5) (2011)
  1919--1932.

\bibitem{gijbels2011Scandinav}
I.~Gijbels, N.~Veraverbeke, M.~Omelka, Estimation of a conditional copula and
  association measures, Scandin. J. Statist. 38 (2011) 766--780.

\bibitem{gijbels2012EJS}
I.~Gijbels, N.~Veraverbeke, M.~Omelka, Multivariate and functional covariates
  and conditional copulas, Electr. J. Statist. 6 (2012) 1273--1306.

\bibitem{gijbels2015EJS}
I.~Gijbels, M.~Omelka, N.~Veraverbeke, Partial and average copulas and
  association measures, Electr. J. Statist. 9 (2015) 2420--2474.

\bibitem{tsai1990truncation}
W.-Y. Tsai, Testing the assumption of independence of truncation time and
  failure time, Biometrika 77~(1) (1990) 169--177.

\bibitem{jondeau2006}
E.~Jondeau, M.~Rockinger, The copula-garch model of conditional dependencies:
  An international stock market application, J. of Internat. Money and Finance
  25 (2006) 827--853.

\bibitem{almeida2017}
C.~Almeida, C.~Czado, Efficient bayesian inference for stochastic time-varying
  copula models, Comput. Statist. Data Anal. 56 (2012) 1511--1527.

\bibitem{soyeung2014regularized}
M.~K. So, C.~Y. Yeung, Vine-copula garch model with dynamic conditional
  dependence, Comput. Statist. Data Anal. 76 (2014) 655--671.

\bibitem{fermanian2015single}
J.-D. Fermanian, O.~Lopez, Single-index copulas, J. Multivariate Anal. 165
  (2018) 27--55.

\bibitem{derumigny2018kendall}
A.~Derumigny, J.-D. Fermanian, About {K}endall's regression, ArXiv preprint,
  arXiv:1802.07613.

\bibitem{derumigny2018kernelBased}
A.~Derumigny, J.-D. Fermanian, {A}bout kernel-based estimation of the
  conditional {K}endall's tau: finite-distance bounds and asymptotic behavior,
  ArXiv preprint, arXiv:1810.06234.

\bibitem{friedman2001elements}
J.~Friedman, T.~Hastie, R.~Tibshirani, The elements of statistical learning,
  Vol.~1, Springer series in statistics New York, 2001.

\bibitem{boyd2011distributed}
S.~Boyd, N.~Parikh, E.~Chu, B.~Peleato, J.~Eckstein, Distributed optimization
  and statistical learning via the alternating direction method of multipliers,
  Foundations and Trends in Machine learning 3~(1) (2011) 1--122.

\bibitem{parikh2014proximal}
N.~Parikh, S.~Boyd, et~al., Proximal algorithms, Foundations and Trends in
  Optimization 1~(3) (2014) 127--239.

\bibitem{scott1992}
D.~Scott, Multivariate Density Estimation: Theory, Practice, and Visualization,
  Wiley, 1992.

\bibitem{wurm2017regularized}
M.~J. Wurm, P.~J. Rathouz, B.~M. Hanlon, Regularized ordinal regression and the
  ordinal{N}et {R} package, ArXiv preprint, arXiv:1706.05003.

\bibitem{ripley2018tree}
B.~Ripley, Tree: classification and regression trees, {R} package version
  1.0-39 (2018).

\bibitem{Breiman1984}
L.~Breiman, J.~Friedman, R.~Olshen, C.~Stone, Classification and Regression
  Trees, Republished by CRC Press., Wadsworth, Belmont, CA., 1984.

\bibitem{lepski1997optimal}
O.~V. Lepski, V.~G. Spokoiny, Optimal pointwise adaptive methods in
  nonparametric estimation, The Annals of Statistics (1997) 2512--2546.

\bibitem{cherubini}
U.~Cherubini, E.~Luciano, W.~Vecchiato, Copula methods in finance, Wiley, 2004.

\bibitem{deheuvels1979}
P.~Deheuvels, La fonction de d\'ependance empirique et ses propri\'et\'es. un
  test non param\'etrique d’ind\'ependance, Acad. Roy. Belg. Bull.Cl. Sci.
  65~(5) (1979) 274--292.

\bibitem{deheuvels1981}
P.~Deheuvels, A kolmogorov–smirnov type test for independence and
  multivariate samples, Rev. Roumaine Math. Pures Appl. 26 (1981) 213--226.

\bibitem{patton2006a}
A.~Patton, Modelling asymmetric exchange rate dependence, Internat. Econom.
  Rev. 47 (2006) 527--556.

\bibitem{patton2006b}
A.~Patton, Estimation of multivariate models for time series of possibly
  different lengths, J. Appl. Econometrics 21 (2006) 147--173.

\bibitem{FermanianWegkamp}
J.-D. Fermanian, M.~Wegkamp, Time-dependent copulas, J. Multivariate Anal. 110
  (2012) 19--29.

\bibitem{geyer1996}
C.~J. Geyer, On the asymptotics of convex stochastic optimization, Tech. rep.,
  Dept. Statistics, Univ. Minnesota (1996).

\bibitem{knight2000asymptotics}
K.~Knight, W.~Fu, Asymptotics for lasso-type estimators, Annals of statistics
  (2000) 1356--1378.

\bibitem{kato2009asymptotics}
K.~Kato, Asymptotics for argmin processes: Convexity arguments, J. Multivariate
  Anal. 100~(8) (2009) 1816--1829.

\end{thebibliography}

\end{document}